\documentclass[]{aa}

\usepackage{graphicx}
\usepackage{amsmath,textcomp}
\usepackage{txfonts}
\usepackage{hyperref}

\def\vect#1{\boldsymbol{#1}}

\usepackage{ulem}
\usepackage{xcolor}

\begin{document} 


\title{Structure of iso-density sets in supersonic isothermal turbulence}

\titlerunning{Structure of iso-density sets}


\author{F. Thiesset\inst{\ref{inst1}}
\and 
C. Federrath\inst{\ref{inst2}}$^,$\inst{\ref{inst3}}} 
\institute{CNRS, CORIA, UMR 6614, Normandy Univ., UNIROUEN, INSA Rouen, 675 Avenue de l'université, BP 12, 76801 Saint Etienne du Rouvray Cedex
\email{fabien.thiesset@cnrs.fr} \label{inst1}
\and
Research School of Astronomy and Astrophysics, Australian National University, Cotter Road, Canberra, ACT 2611, Australia \label{inst2}
\and 
Australian Research Council Centre of Excellence in All Sky Astrophysics (ASTRO3D), Cotter Road, Canberra, ACT 2611, Australia  \\
\email{christoph.federrath@anu.edu.au}\label{inst3} 
}

\authorrunning{F. Thiesset \and C. Federrath}

\abstract
  {The gas density structure of the cold molecular phase of the interstellar medium is the main controller of star formation.}
  {A theoretical framework is proposed to describe the structural content of the density field in isothermal supersonic turbulence.}
  {It makes use of correlation and structure functions of the phase indicator field defined for different iso-density values. The relations between these two-point statistics and the geometrical features of iso-density sets such as the volume fraction, the surface density, the curvature, and fractal characteristics are provided. An exact scale-by-scale budget equation is further derived revealing the role of the turbulent cascade and dilation on the structural evolution of the density field. Although applicable to many flow situations, this tool is here first invoked for characterising supersonic isothermal turbulence, using data from the currently best-resolved numerical simulation. }
  {We show that iso-density sets are surface fractals rather than mass fractals, with dimensions that markedly differ between dilute, neutral, and dense regions. The surface--size relation is established for different iso-density values. We further find that the turbulent cascade of iso-density sets is directed from large towards smaller scales, in agreement with the classical picture that turbulence acts to concentrate more surface into smaller volumes. Intriguingly, there is no range of scales that complies with a constant transfer rate in the cascade, challenging our fundamental understanding of interstellar turbulence. Finally, we recast the virial theorem in a new formulation drawing an explicit relation between the aforementioned geometrical measures and the dynamics of iso-density sets. }
  {}

\keywords{ISM: kinematics and dynamics, Hydrodynamics, Turbulence, Methods: analytical, numerical, statistical}


\maketitle
\section{Introduction} \label{sec:intro}

Turbulence is one of the key processes that shapes the spatial and temporal evolution of matter and energy across nearly all scales, from the laboratory up to astrophysical scales. When associated with the interstellar medium (ISM), turbulence is observed to lie in the supersonic regime \citep[see for example][and references therein]{Elmegreen2004,MacLow2004,McKee2007,Hennebelle2012}, yielding coupled correlations between velocity and density fluctuations at all scales. Therefore, a physical model for predicting the spatial organisation of the gas density (and its tracers) throughout the ISM needs to account for the interactions between the velocity and density fields. This constitutes one of the open challenges in the astrophysics community and has key relevance for understanding the structure and dynamics of the ISM, and in particular the physical conditions for the formation of stars \citep{PadoanEtAl2014}.

There has been significant progress in the statistical characterisation of density fluctuations inferred from either observations or numerical simulations of the ISM. One of the most popular statistical tool is the one-point probability density function as it comes as input in several star formation models \citep{Padoan2002,Krumholz2005,Hennebelle2008,Padoan2011,Federrath2012,BurkhartMocz2019,AppelEtAl2022}. It has been shown that for an isothermal gas in supersonic turbulence, the volume- and mass-weighted density fluctuations comply relatively well with a log-normal distribution. The latter arises naturally by assuming a random multiplicative process and the application of the central limit theorem for the density evolution \citep{Vazquez-Semadeni1994,Passot1998,Kritsuk2007,FederrathKlessenSchmidt2008,Federrath2010}. Significant deviations from the log-normal distribution are however observed when gravity \citep[see for instance][]{Kritsuk2011,Federrath2013,GirichidisEtAl2014,Khullar2021}, magnetic fields, and/or stellar feedback \citep[see for instance][]{Krumholz2012,Myers2014,KainulainenFederrathHenning2014,Federrath2015,SchneiderEtAl2016} are included.

More insights into the spatial distribution of matter in the ISM can be provided by probing the microstructure of the density field. By microstructure, we refer here to any scale-dependent features of the density field and its tracers \citep[the reader may refer to][for a complete review of several micro-structural descriptors]{Elmegreen2004}. Such analysis can be carried out using two-point statistics: for example, correlation functions, structure functions, Fourier spectra or $\Delta$-variance techniques, or principal component analysis \citep[see for instance][]{Stutzki1998,OssenkopfMacLow2002,Padoan2004,HeyerBrunt2004,Kim2005,Kritsuk2006,Federrath2010,RomanDuvalEtAl2010}. The micro-structural content of the density field can also be assessed using geometrical approaches, which include fractal techniques \citep{Elmegreen1996,Stutzki1998,Federrath2009,Audit2010,Kritsuk2007,BeattieFederrathKlessen2019,BeattieEtAl2019b} or multi-fractal spectra \citep{Chappell2001}. Independent of the tool, one generally seeks to find some power-law variations of the observable with respect to the scale. This power-law behaviour is very useful to spark phenomenological scenarios that describe the physics at play in the structural evolution of the gas density in the ISM. However, the existence of some power laws is generally observed, sometimes predicted using dimensional arguments, but it is not generally derived from first principles. One attempt to fill this gap is presented by \citet{Galtier2011} and \citet{Ferrand2020} who derived the exact generalised Kolmogorov equation directly from the compressible Navier-Stokes equations. The work by \cite{Aluie_2013} also provides some theoretical insights into the scale distribution of compressible turbulence based on the coarse grained Navier-Stokes equations.  Nevertheless, such exact scale-by-scale budget equations, although extremely valuable to describe the physics at play, still require some closures for being used as a predictive tool.

In the present study, we aim to provide new insights into the role of supersonic turbulence in determining the spatial structure of the gas density in the ISM. We propose using a two-point statistical analysis of the phase indicator field defined from different density thresholds. Such an approach is encountered in other branches of physics, dealing with for example, heterogeneous materials \citep{Adler1990, Torquato2002, Teubner1990, Kirste1962, Frisch1963, Berryman1987} or fractal aggregates \citep{Sorensen2001,Moran2019}. Its application to fluid mechanics is relatively scarce, although it has been applied with success in single-phase turbulence \citep{Hentschel1984,Vassilicos1991,Vassilicos1992,Vassilicos1996,Elsas2018,Gauding2022} and multiphase turbulent flows \citep{Lu2018,Lu2019,Thiesset2020,Thiesset2021}. There are several theoretical results based on robust mathematical grounds allowing the two-point statistics of iso-value sets to be related to some geometrical and/or fractal properties. An exact transport equation for such two-point statistics was further derived \citep[][]{Thiesset2020,Gauding2022}, which makes the interactions explicit between the probed field variable and the turbulent velocity field. It is therefore believed that this tool could be promising to investigate the density fluctuations in supersonic turbulence. Although virtually applicable to all scenarios involving turbulent flows, this framework was first appraised using data from a high-resolution simulation of supersonic isothermal turbulence \citep{Federrath2021}. 

The rest of the paper is organised as follows. Section~\ref{sec:theory} gathers the main theoretical derivations. The numerical database and post-processing procedures are introduced in Section~\ref{sec:numerics}. Our results are presented in Section~\ref{sec:results} and conclusions are drawn in Section~\ref{sec:conclusion}.

\section{A structural descriptor of iso-density sets} \label{sec:theory}

\subsection{The phase indicator field}
The analysis is based on the phase indicator function $\phi(\vect{x},t)$ defined as:
\begin{eqnarray}
    \phi(\vect{x},t)=\begin{cases}
                1 ~~ {\rm when ~\rho(\vect{x},t)>\rho_{\rm th}}  \\
                0 ~~ {\rm otherwise}. 
             \end{cases} \label{eq:phi}
\end{eqnarray}
This quantity reads as the probability that the density $\rho$ is larger than a certain threshold $\rho_{\rm th}$ at a given position in space $\vect{x}$ and time $t$. It is also sometimes referred to as the excursion set or here iso-density set. The phase indicator function was introduced notably to characterise heterogeneous media such as composite materials and/or porous media \citep[see for instance][]{Debye1957,Porod1951}. Such fields are discontinuous by nature with two (or more) phases separated by an interface. In supersonic turbulence, the presence of shocks may lead also to local discontinuities of the density field. Despite, one can always define a phase indicator field, in the presence or absence of discontinuities, and hence it is applicable here to the density field, even in presence of shocks.

Investigating the properties of $\phi(\vect{x},t)$ for different iso-values $\rho_{\rm th}$ allows the geometry of the density field to be characterised. Low and high values of $\rho_{\rm th}$ correspond the dilute and dense regions, respectively. The relevant geometrical properties of $\phi$ investigated here are detailed below.

The first quantity is the volume fraction, which is simply defined as the ratio between the phase indicator volume and the averaging volume:
\begin{equation}
  \langle \phi \rangle = \frac{1}{V} \int \phi(\vect{x},t) {\rm d} V. \label{eq:volume_fraction}
\end{equation} 
The brackets in Eq. \eqref{eq:volume_fraction} denote a volume average over the volume $V$. The volume fraction $\langle \phi \rangle$ has no units and is comprised between 0 and 1.

The second relavant geometrical feature of the field $\phi$ is the surface density, which represents the surface area of the interface separating $\phi = 1$ and $\phi=0$, divided by the averaging volume: 
  \begin{equation}
  \Sigma = \langle |\vect{\nabla} \phi(\vect{x},t)| \rangle = \frac{1}{V} \int |\vect{\nabla} \phi(\vect{x},t)| {\rm d} V. \label{eq:sigma_def}
  \end{equation}
We note that in the astrophysics community, the surface density generally refers to the cloud mass divided by the cloud bounding surface area. It is thus given in units of mass per area. Let this quantity be noted $\Sigma_m$. In Eq. \eqref{eq:sigma_def}, what we call surface density $\Sigma$, is a purely geometrical quantity with no connection to the mass, defined by the area of the iso-density surface divided by the averaging volume. Hence, it is in units of inverse of length. Consequently, for a cloud with fixed mass, if its bounding surface increases, the mass surface density $\Sigma_m$ decreases while, the geometrical surface density $\Sigma$ increases. Therefore, they evolve in opposite directions. One possible way to relate these two quantities is to compute the mass fraction $\langle \rho \phi \rangle$ (that is the mass contained in the iso-volume defined by $\rho(\vect{x},t)>\rho_{\rm th}$ divided by the averaging volume $V$) and then one has $\Sigma_m = \langle \rho \phi \rangle/\Sigma$.

We can go beyond and compute the statistics of the spatial increment for $\phi$, which is written $\delta \phi$ and is defined as the difference of $\phi$ between two points $\vect{x} + \vect{r}$ and $\vect{x}$, arbitrarily separated in space by a distance $\vec{r}$:
\begin{eqnarray}
    \delta \phi (\vect{r}, t) = \phi(\vect{x} + \vect{r}, t) - \phi(\vect{x}, t) \label{eq:increments_pm}.
\end{eqnarray}
We consider in particular the second-order moment of $\delta \phi$, also called the second-order structure function, defined by:
\begin{eqnarray}
    \langle  (\delta \phi )^2 \rangle = \frac{1}{V} \int (\delta \phi)^2 {\rm d}V.
\end{eqnarray}

The phase indicator function can also be studied through its two-point correlation function defined as:
\begin{eqnarray}
  \langle  \phi_m^+ \phi_m \rangle = \frac{1}{V} \int \phi_m(\vect{x}+\vect{r}) \phi_m(\vect{x}) {\rm d}V. \label{eq:correlation_def}
\end{eqnarray}
The + superscript in Eq. \eqref{eq:correlation_def} is used to denote that the quantity is taken at point $\vect{x}+\vect{r}$ while $\phi_m$ denotes the phase indicator for the minority phase at point $\vect{x}$. The minority phase is defined by
\begin{eqnarray}
  \phi_m(\vect{x},t)=\begin{cases}
    \phi(\vect{x},t) ~~~~~~~~~~ {\rm if ~} \langle \phi \rangle \leq 0.5  \\
    1-\phi(\vect{x},t) ~~~~ {\rm if ~} \langle \phi \rangle > 0.5. 
  \end{cases}
\end{eqnarray}
We thus have $\langle \phi_m \rangle = \min(\langle \phi \rangle, 1-\langle \phi \rangle)$. The majority phase is defined as the complementary set of the minority phase and is equal to $\phi^\prime = 1-\phi_m$. Contrary to the autocorrelation function, the second-order structure function is the same when computed from the majority or minority phase.
  
\cite{Thiesset2020} derived the relation between the correlation function and the second-order structure function:
\begin{eqnarray}
  \frac{\langle  \phi_m^+ \phi_m \rangle}{\langle  \phi_m \rangle} = 1 - \frac{\langle  (\delta \phi )^2 \rangle}{2 \langle \phi_m \rangle}. \label{eq:correl-struct}
\end{eqnarray}
Although Eq. \eqref{eq:correl-struct} reveals that the correlation and structure function are intimately linked, we subsequently show later in this paper that they provide different information about the system. 

In general (inhomogeneous and anisotropic) situations, both the second-order structure function $\langle  (\delta \phi )^2 \rangle$ and the correlation function $\langle  \phi_m^+ \phi_m \rangle$ depend on the 3-dimensional separation vector $\vect{r}$ and the 3-dimensional position vector $\vect{x}$. Homogeneity can be invoked such that two-point statistics are invariant by translation, thereby dropping the dependence to the vector $\vect{x}$. In case of isotropic media, two-point statistics depend only on the modulus of the separation vector $r \equiv |\vect{r}|$. The numerical data detailed in section \ref{sec:numerics} and discussed in section \ref{sec:results} correspond to homogeneous and isotropic turbulence, which means that the two-point statistics discussed hereafter are function of $r$, only. 

The framework presented here was used in its fully inhomogeneous and anisotropic version in \cite{Thiesset2021}. These authors showed that the relations to be discussed below apply to the 'homogeneised' (after application of a spatial average) and 'isotropised' (after application of an angular average over all orientations of the separation vector) version of the original inhomogeneous and anisotropic media.

\subsection{Asymptotic behaviour of two-point statistics}

The asymptotic behaviour of $\langle (\delta \phi )^2 \rangle$ and $\langle  \phi_m^+ \phi_m \rangle$ for different range of scales are known and are detailed below.

\subsubsection{At small scales} Since $\phi$ can take only 1 or 0 values, we have $\phi^2 \equiv \phi$, and hence the correlation function $\langle  \phi_m^+ \phi_m \rangle$ at $r=0$ is given by
\begin{eqnarray}
  \langle  \phi_m^+ \phi_m \rangle (r=0) = \langle  \phi_m \rangle. 
\end{eqnarray}
Therefore, the correlation function at $r=0$ gives information about the volume fraction of the minority phase $\langle \phi_m \rangle$. The second-order structure function $\langle (\delta \phi )^2 \rangle$ is 0 at $r=0$. 

The asymptotic regime when the separation $r$ tends to zero was derived by \citet{Porod1951}, \citet{Guinier1955}, and \citet{Debye1957}, who proved that for homogeneous isotropic media: 
\begin{eqnarray}
  \lim_{r \to 0} \langle  \phi_m^+ \phi_m \rangle = \langle \phi_m \rangle - \frac{\Sigma r}{4}. \label{eq:corr_small_scale}
\end{eqnarray}
This implies for $\langle  (\delta \phi )^2 \rangle$ the following limit:
\begin{eqnarray}
  \lim_{r \to 0} \langle  (\delta \phi )^2 \rangle = \frac{\Sigma r}{2}. \label{eq:dphi2_small_scale}
\end{eqnarray}
Eqs.~\eqref{eq:corr_small_scale} and \eqref{eq:dphi2_small_scale} can be seen as the 3D extension of the classical Buffon needle problem. It shows that if the interface between $\phi = 1$ and $\phi=0$ becomes planar when observed at sufficiently small scales, the probability that the two points $\vect{x}$ and $\vect{x} + \vect{r}$ lie on each side of the interface is then simply proportional to the surface density $\Sigma$ and the distance $|\vect{r}|$ between the two points. Eqs.~\eqref{eq:corr_small_scale} and \eqref{eq:dphi2_small_scale} remain valid in anisotropic media when two-point statistics are angularly averaged over all orientations of the vector $\vect{r}$ \citep{Berryman1987}. This linear regime is observed only if the interface is planar at some resolution scales. In case of purely fractal sets, revealing rough interfaces at all scales, this regime is not likely to be observed.

\subsubsection{At larger, yet small scales} For slightly larger values of the separation $r$, the curvature of the interface becomes perceptible, and one needs to account for the next terms in the small-scale expansion of $\langle  \phi_m^+ \phi_m \rangle$ and $\langle  (\delta \phi )^2 \rangle$. These were first derived by \citet{Kirste1962} and \citet{Frisch1963}, and later by \citet{Teubner1990} and \citet{Ciccariello1995}, and read:
\begin{eqnarray}
  \lim_{r \to 0} \langle  \phi_m^+ \phi_m \rangle = \langle \phi_m\rangle - \frac{\Sigma r}{4} \left( 1 - \langle C \rangle_s\frac{r^2}{8} \right). \label{eq:H_G_expansion_corr}
\end{eqnarray}
By virtue of Eq.~\eqref{eq:correl-struct}, we then have for $\langle  (\delta \phi )^2 \rangle$:
\begin{eqnarray}
  \lim_{r \to 0} \langle  (\delta \phi )^2 \rangle = \frac{\Sigma r}{2} \left( 1 - \langle C \rangle_s\frac{r^2}{8} \right). \label{eq:H_G_expansion}
\end{eqnarray}
The quantity $\langle C\rangle_s$ in Eqs.~\eqref{eq:H_G_expansion_corr} and \eqref{eq:H_G_expansion} is related to the mean and Gaussian curvature, denoted $H$ and $G$:
\begin{eqnarray}
  \langle C\rangle_s = \left[\langle H^2\rangle_s - \frac{\langle G\rangle_s}{3}\right],
\end{eqnarray}
where $\langle \bullet \rangle_s$ denotes a surface-area-weighted average: $\langle \bullet \rangle_s \Sigma = \langle \bullet |\vect{\nabla} \phi| \rangle$. Eqs.~\eqref{eq:H_G_expansion_corr} and \eqref{eq:H_G_expansion} reveal that the quantity $1/\langle C\rangle_s^{1/2}$ can be associated with the transition scale where the interface starts being curved. In anisotropic media, Eq.~\eqref{eq:H_G_expansion} was shown to hold true when an angular average is applied on two-point statistics \citep{Thiesset2021}.

\subsubsection{In the intermediate range of scales}
The correlation and structure functions are known to provide information about the fractal features of the object, if any. As stated for example by \cite{Sreenivasan1989}, an object is likely to exhibit a fractal scaling in a range of scales lying between an inner cutoff $\eta_i$ (here a scale somehow related to $\langle C \rangle_s^{-1/2}$) and an outer cutoff $\eta_o$ (a kind of integral length-scale). If the separation between $\eta_i$ and $\eta_o$ is sufficiently large, then one should expect $\langle \phi_m^+ \phi_m \rangle$ and/or $\langle (\delta \phi )^2 \rangle$ to follow a power law with an exponent that can be related to the fractal dimension. 

\begin{figure}
  \centering
  \includegraphics[width=\linewidth]{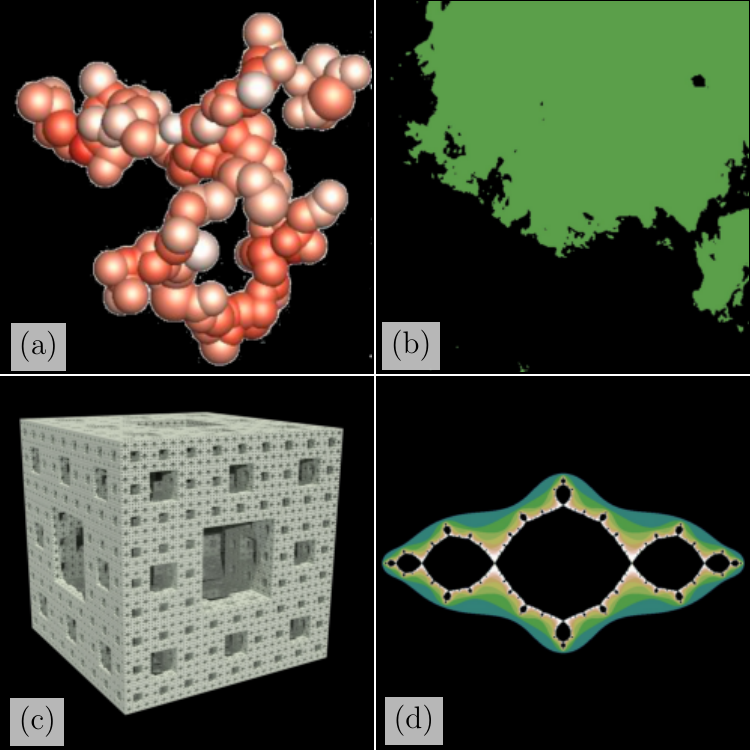}
  \caption{Illustrations of different categories of fractals. Examples of mass-fractals: (a) a soot particle and (c) the Menger sponge.  Examples of surface-fractals : (b) a coastline and (d) the Julia set.} \label{fig:fractals}
\end{figure}

However, in this context, it is important to make the distinction between what is referred to as a mass-fractal and a surface-fractal. Schematically, a mass-fractal is an object whose bulk reveals some fractal features while its surface remains smooth. Soot aggregates issued from combustion of hydrocarbon fuels typically fall in this category \citep{Sorensen2001,Moran2019}. Such an object is presented in Fig.~\ref{fig:fractals} (a). It is composed of a set of small spherical particles with constant density whose surface is smooth, and assembles to form an object with fractal features. Another example is the Menger sponge as illustrated in Fig.~\ref{fig:fractals} (c). The latter is composed of perforated cubes with scale-similarity. Except at the corners, its surface is always planar.

A surface-fractal is the opposite: an object whose body remains compact while its surface is fractal. Some examples are given in the right column of Fig.~\ref{fig:fractals}. Fig.~\ref{fig:fractals}(b) shows a coastline, and the Julia set is illustrated in Fig.~\ref{fig:fractals}(d), which are examples of typical surface-fractals.

This distinction is of major importance here since the correlation and structure function for mass- and surface-fractals exhibit different behaviour in the intermediate range of scales. Indeed, for instance \citet{Sorensen2001} and \citet{Wong1992} (and references therein) mentioned that for mass-fractals:
\begin{eqnarray}
  \langle \phi_m^+ \phi_m \rangle \sim r^{\xi_m},
\end{eqnarray}
where $\xi_m = D_m - 3$ with $D_m$ the mass-fractal dimension.

By contrast, for surface-fractals \citep{Wong1992} with dimension $D_s$:
\begin{eqnarray}
  \langle \phi_m^+ \phi_m \rangle  = \langle \phi_m \rangle - K r^{\xi_s}, \label{eq:corr_surface_fractal}
\end{eqnarray}
where $\xi_s = 3- D_s$, and $K$ is a constant. Eq.~\eqref{eq:corr_surface_fractal} together with Eq.~\eqref{eq:correl-struct} indicates that: 
\begin{eqnarray}
  \langle (\delta \phi)^2 \rangle \sim r^{\xi_s}. \label{eq:strufu_surface_fractal}
\end{eqnarray}
Therefore, probing the scale dependence of $\langle \phi_m^+ \phi_m \rangle$ and $\langle (\delta \phi)^2 \rangle$ separately enables \textit{(i)} to assess whether the object under consideration is rather a mass- or surface-fractal, and \textit{(ii)} to estimate the corresponding dimension $D_m$ or $D_s$.

\begin{figure}
  \includegraphics[width=\linewidth]{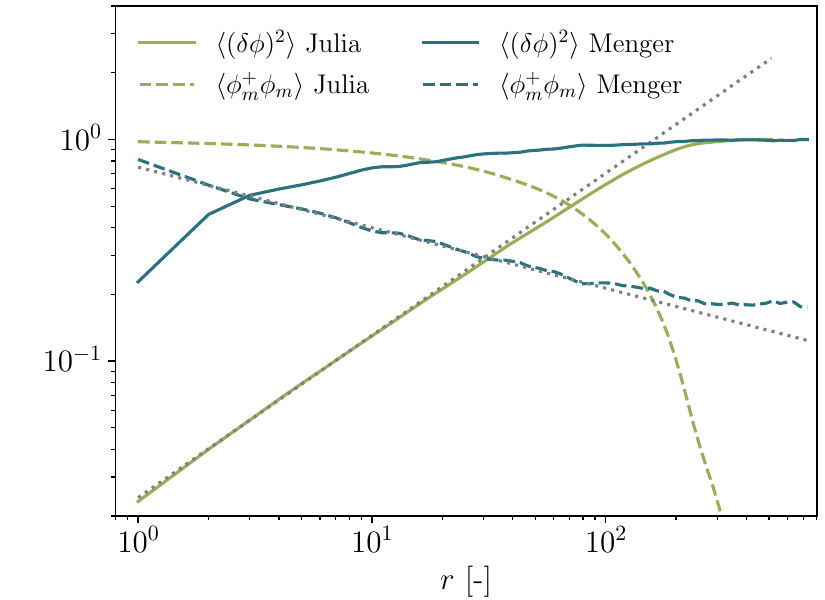} 
  \caption{Correlation and structure function for the Menger sponge and the axisymmetric Julia set ($z \rightleftharpoons z^2 - 1$). The grey dotted line represents the theoretical power laws with $D_m = \log_3(20)$ for the Menger sponge and $D_s \approx 2.27$ for the Julia set.} \label{fig:Menger_Julia}
\end{figure}

An example is given in Fig.~\ref{fig:Menger_Julia} where we have computed $\langle \phi_m^+ \phi_m \rangle$ and $\langle (\delta \phi)^2 \rangle$ for the Menger sponge (a mass-fractal) and the Julia set (a surface-fractal). For the Julia set, $\langle (\delta \phi)^2 \rangle$ reveals an appreciable power-law range, while no such behaviour is observed for $\langle \phi_m^+ \phi_m \rangle$. The opposite is observed for the Menger sponge. In each situation the computed power-law exponent complies well with the theoretical values of $D_m = \log_3(20)\approx2.73$ and $D_s \approx 2.27$ for the Menger sponge and the Julia set, respectively. It is worth noting that in Fig.~\ref{fig:Menger_Julia}, $\langle (\delta \phi)^2 \rangle$ does not reveal any range of scales complying with a linear regime with respect to the separation $r$ (Eq. \eqref{eq:dphi2_small_scale}). This means that the Julia and Menger sets are rough at all scales.

The distinction between mass- and surface-fractal yields important consequences. In particular, the mass--size distribution for mass-fractals is 
\begin{eqnarray}
  M(r) \sim r^{D_m}, \label{eq:mass_mass_fractal}
\end{eqnarray}
while for surface fractals,
\begin{eqnarray}
  M(r) \sim r^3.  \label{eq:mass_surface_fractal}
\end{eqnarray}
For a surface-fractal, the surface area measured at scale $r$ (known as the surface-scale distribution) is given by \citep{Sreenivasan1989,Wong1992}
\begin{eqnarray}
  S(r) \sim r^{D_s-2}.  \label{eq:surface_surface_fractal}
\end{eqnarray}
One can further imagine a situation where a mass-fractal is bounded by a surface-fractal. In this case, \citet{Wong1992} obtained that:
\begin{eqnarray}
  M(r) \sim r^{D_m} \left(1-A\left(\frac{r}{R}\right)^{3-D_s}\right)  \label{eq:mass_mass_surface_fractal}
\end{eqnarray}
for any $r < R$ where $R^3$ is the volume enclosed by the surface given by $\langle \phi_m \rangle  = R^3/V$, and $A$ is a constant of order unity. It is noted that the correction $A(r/R)^{3-D_s}$ is perceptible only at scales close to $R$.

\subsubsection{At large scales} The asymptotic limit of the correlation function at large scales is
\begin{eqnarray}
\lim_{r \to \infty} \langle \phi_m^+ \phi_m \rangle = \langle \phi_m \rangle^2.
\end{eqnarray}
This result implies that for the second-order structure function 
\citep{Thiesset2020,Thiesset2021,Gauding2022}:
\begin{eqnarray}
  \lim_{r \to \infty} \langle  (\delta \phi )^2 \rangle = 2 \langle \phi \rangle (1-\langle \phi \rangle) \label{eq:dphi2_large_scales}
\end{eqnarray}
Hence, in the limit of large scales, both $\langle \phi_m^+ \phi_m \rangle$ and $\langle  (\delta \phi )^2 \rangle$ give information about the volume fraction.

\subsubsection{Summary of the schematic representation of $\phi$}

\begin{figure*}
  \includegraphics[width=\linewidth]{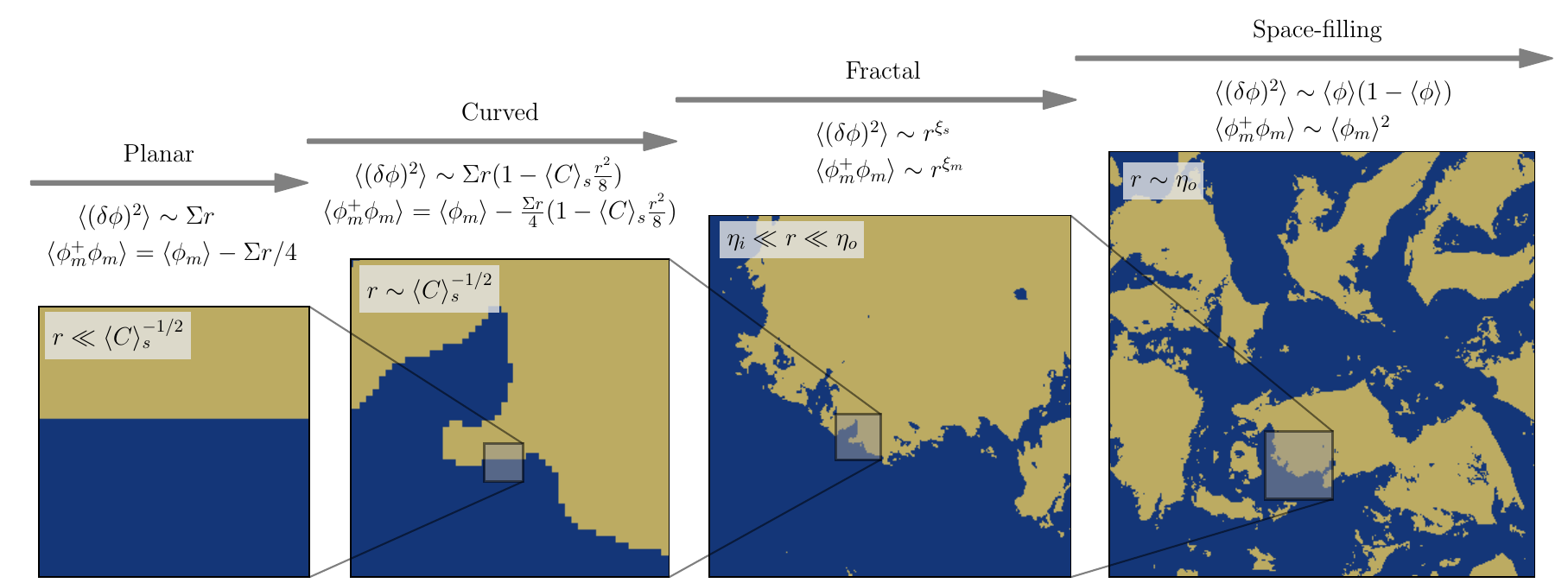}  
  \caption{Schematic representation of the different asymptotic regimes of $\langle (\delta \phi)^2\rangle$ and $\langle  \phi_m^+ \phi_m \rangle$ for different scale ranges from small to large scales.} \label{fig:dphi2_asymptotics}
\end{figure*}

These different asymptotic regimes are schematically summarised in Fig.~\ref{fig:dphi2_asymptotics}. This figure shows that when probed at asymptotically small scales, the interface seems planar. In this regime, $\langle  (\delta \phi )^2 \rangle$ and $\langle  \phi_m^+ \phi_m \rangle$ are intimately linked to the surface density $\Sigma$. At slightly larger scales, that is for scales $r \sim \langle C\rangle^{-1/2}$, the interface curvature starts to become visible. Both $\langle  (\delta \phi )^2 \rangle$ and $\langle  \phi_m^+ \phi_m \rangle$ are then given by Eqs. \eqref{eq:H_G_expansion} and \eqref{eq:H_G_expansion_corr}, respectively. If $r$ lies well between $\eta_i$ and $\eta_o$, then a fractal scaling can possibly be observed. In this situation $\langle  (\delta \phi )^2 \rangle$ follows a power law with an exponent $\xi_s = 3-D_s$ if the object is surface-fractal. Conversely, if the object is mass-fractal, then the correlation function exhibits a power law with exponent $\xi_m = D_m-3$. Finally, at large scales, $\langle  (\delta \phi )^2 \rangle$ starts being a volumetric descriptor and reaches the asymptotic value of $2\langle \phi \rangle (1-\langle \phi \rangle)$, while $\langle  \phi_m^+ \phi_m \rangle$ tends towards $\langle \phi_m \rangle^2$.

\begin{figure*}
  \includegraphics[width=\linewidth]{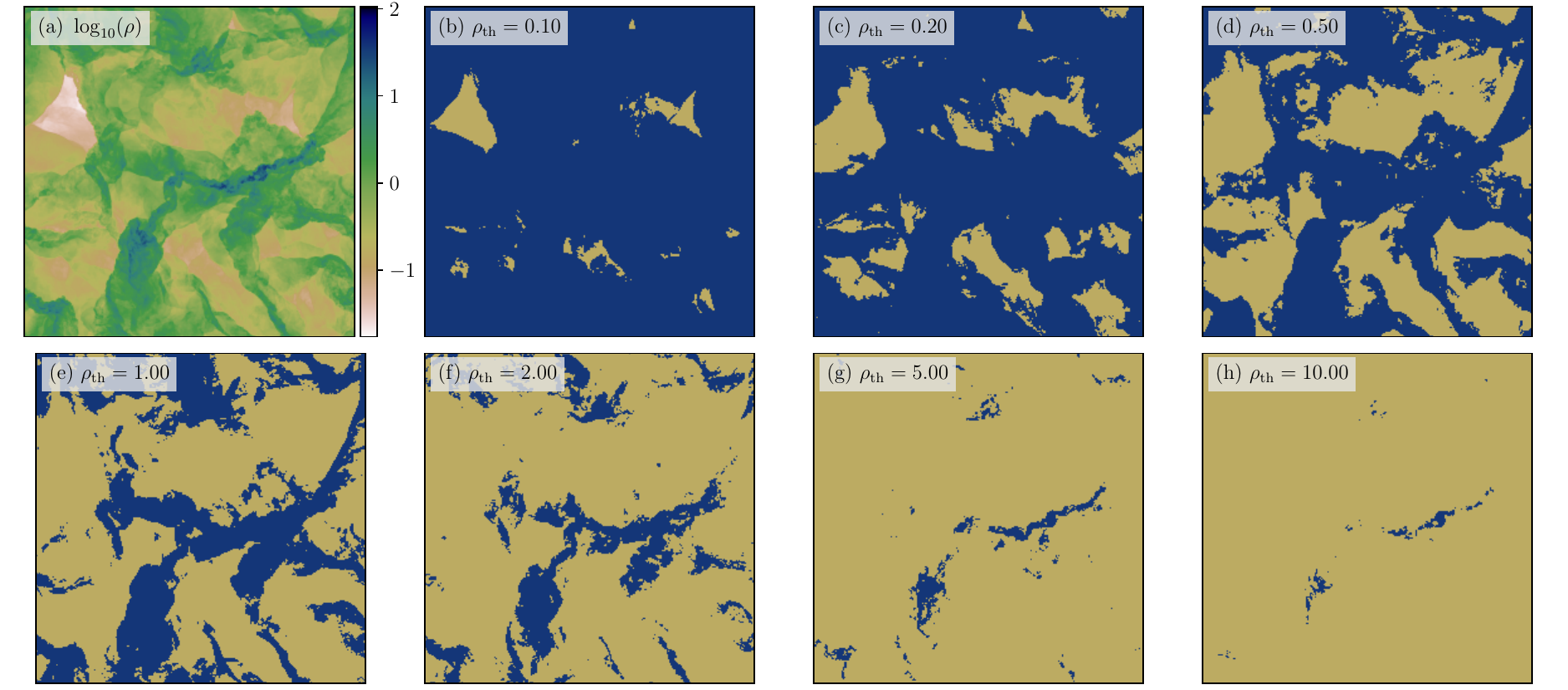}  
  \caption{Two-dimensional slices of the density field and their associated phase indicator field. (a) slice of $\log(\rho/\rho_0)$ with light (dark) colours corresponding to low (high) values. (b-h) Corresponding iso-density set with $\rho_{\rm th}=0.1$ to 10.0, as indicated in the legend of each panel. The yellow and blue regions correspond to $\phi=0$ and $\phi=1$, respectively.} \label{fig:slice}
\end{figure*}

As an overall conclusion, the quantities $\langle  (\delta \phi )^2 \rangle$ and $\langle  \phi_m^+ \phi_m \rangle$ contain information about the surface density, the interface curvature, the fractal characteristics and the volume fraction. All these quantities are important geometric measures that are contained in and characterised by the rather simple structural descriptor $\phi$ defined in Eq.~(\ref{eq:phi}).

\subsection{Scale-by-scale budget}
The analysis based on the correlation and structure functions of the phase indicator $\phi$ can be supplemented by a transport equation, which is known as a scale-by-scale budget. When applied to the density field (a conserved quantity), the time and space evolution of $\phi$ is given by \citep{Thiesset2020,Thiesset2021,Gauding2022},
\begin{eqnarray}
    \partial_t \phi + \vect{u} \cdot \vect{\nabla} \phi = 0, \label{eq:phi_eq}
\end{eqnarray}
where $\vect{u}$ is the fluid velocity at the interface. Using the machinery described by \citet{Thiesset2020} and \citet{Gauding2022}, one can derive the transport equation for $\langle (\delta \phi)^2 \rangle$ (and $\langle \phi_m^+ \phi_m \rangle$). The general scale-by-scale budget for reacting and diffusive quantities evolving in non-stationary, inhomogeneous, anisotropic, possibly compressible flows is provided by \cite{Gauding2022}. Here we provide the formulation for a statistically homogeneous and stationary flow, as in our numerical simulations. In this case, the equation for $\langle (\delta \phi)^2\rangle$ simplifies to
  \begin{eqnarray}
    \vect{\nabla_r} \cdot \langle (\delta \vect{u}) (\delta \phi)^2\rangle = 2\langle (\vect{\nabla} \cdot \vect{u})^\oplus (\delta \phi)^2 \rangle.
 \label{eq:dphi2_eq_div}
\end{eqnarray}
The quantity $\bullet^\oplus = (\bullet^+ + \bullet)/2$ is the arithmetic mean of the quantity $\bullet$ between the points $\vect{x}$ and $\vect{x+r}$. The transport equation for the correlation function was obtained by \cite{Gauding2022}. Eq. \eqref{eq:dphi2_eq_div} was derived by assuming that the velocity $\vect{u}$ is differentiable when crossing the interface. Hence, this formulation of scale-by-scale is restricted to cases where the velocity can be assumed to be smoothly varying on each side of the iso-density surface. In case of strong shocks, associated with discontinuous velocity jumps, the weak formulations of the scale-by-scale budgets, such as those discussed by \cite{Duchon2000,Saw2016,Galtier2018,Dubrulle2019} for the turbulent kinetic energy, should be derived and used instead.

By further taking advantage of isotropy, the transfer term on the left-hand side of Eq.~\eqref{eq:dphi2_eq_div}, which writes as the divergence in scale-space of the flux $\langle (\delta \vect{u}) (\delta \phi)^2\rangle$, can be expressed in spherical coordinates, which leads to:
\begin{eqnarray}
  \langle (\delta u_\parallel) (\delta \phi)^2\rangle = \frac{2}{r^2} \int_0^r r^2 \langle (\vect{\nabla} \cdot \vect{u})^\oplus (\delta \phi)^2 \rangle {\rm d} r.  \label{eq:dphi2_eq}
\end{eqnarray}
The quantity $u_\parallel = \vect{u} \cdot \vect{r}/r$ is the longitudinal velocity: the velocity component in the direction of $\vect{r}$. 

Equation \eqref{eq:dphi2_eq} is exact and is derived without any other hypothesis than the one invoked above (homogeneous, isotropic, stationary fields). It reveals the effect of velocity and velocity divergence on the evolution of the microstructure of the density field as measured through $\langle (\delta \phi)^2\rangle$ or $\langle \phi_m^+ \phi_m \rangle$ at a given threshold $\rho_{\rm th}$. The term on the left-hand side of Eq.~\eqref{eq:dphi2_eq} is the scale-by-scale transport of the quantity $\phi$ by the velocity field. In the classical \cite{Kolmogorov1941} or \cite{Yaglom1949} theory, for incompressible turbulence and turbulent mixing, respectively \citep[see also][]{Danaila2004}, this term is generally associated with the cascade process \citep[see also][for more recent derivations of a Kolmogorov-type theory of compressible isothermal turbulence]{Galtier2011,Ferrand2020}. When negative, the flux of $(\delta \phi)^2$ is directed towards small scales (a direct cascade), while positive $\langle (\delta u_\parallel) (\delta \phi)^2\rangle>0$ is referred to as an inverse cascade from small to large scales.

The process on the right-hand side of Eq.~\eqref{eq:dphi2_eq} arises in flows with non-zero velocity divergence. When positive (negative), this process acts in expanding (contracting) the microstructure in the space of separation $r$. The velocity divergence is thus a sink/source term in the scale-by-scale budget of the iso-density field, which counteracts the transfer process. Very similar conclusions were drawn by \citet{Galtier2011} and \citet{Ferrand2020}, who showed that the divergence of the velocity acts similarly in the scale-by-scale budget for the velocity structure functions. 

The asymptotic behaviour of the different terms in Eq.~\eqref{eq:dphi2_eq} at small scales was determined by \citet{Gauding2022}. For the flux term, the limit is 
\begin{eqnarray}
  \lim_{r\to 0} - \langle (\delta u_\parallel) (\delta \phi)^2\rangle = \mathbb{K}\Sigma \frac{r^2}{8}, \label{eq:dudphi2_small_scales}
\end{eqnarray}
where $\mathbb{K} = \langle - \vect{n} \cdot \vect{\nabla}\vect{u} \cdot \vect{n} \rangle_s$ is here the tangential component of the strain rate acting on the iso-surface $\rho(\vect{x},t)=\rho_{\rm th}$ \citep{Candel1990}. The vector $\vect{n}=-\vect{\nabla} \phi / |\vect{\nabla} \phi|$ is the unit normal vector to the interface. The other component of the strain rate writes $\langle \vect{\nabla} \cdot \vect{u}\rangle_s $ and is due to the velocity divergence. This term arises from the limit of the term on the right-hand side of Eq.~\eqref{eq:dphi2_eq}, in the limit of small separations: 
\begin{eqnarray}
  \frac{2}{r^2} \int_0^r r^2 \langle (\vect{\nabla} \cdot \vect{u})^\oplus (\delta \phi)^2 \rangle {\rm d} r = \langle \vect{\nabla} \cdot \vect{u}\rangle_s \Sigma \frac{r^2}{8}.
\end{eqnarray}
The turbulent flow that we subsequently investigate in the following, is at steady state. In this situation, the sum of the two components of the strain rate is zero. 

\section{Numerical setup and post-processing} \label{sec:numerics}

Here we analyse data from a highly resolved numerical simulation of supersonic isothermal turbulence. The database and numerical methods used in this simulation are described in detail in \citet{Federrath2021} \citep[see also][]{Ferrand2020}. We briefly summarised the main methods below.

The compressible Euler equation in three dimensions:
\begin{eqnarray}
  \partial_t \rho \vect{u} + \vect{\nabla} \cdot \rho \vect{u} \otimes \vect{u} = -\vect{\nabla} p + \rho \vect{F}, \label{eq:Euler}
\end{eqnarray}
is solved, together with the continuity equation
\begin{eqnarray}
  \partial_t \rho  + \vect{\nabla} \cdot \rho \vect{u} = 0,
\end{eqnarray}
using the code FLASH \citep{Fryxell2000,DubeyEtAl2008}. These equations are solved on a triply periodic Cartesian mesh using a positivity-preserving MUSCL-Hancock HLL5R Riemann scheme \citep{Waagan2011}. An isothermal equation of state for a perfect gas, $p = \rho c_s^2$, is used to relate the pressure $p$ and the density $\rho$ through the sound speed $c_s$, which is constant. Turbulence statistics are maintained at steady state using a forcing $\rho\vect{F}$. The latter acts at large scales and is composed of a half solenoidal and half compressive mode, termed natural mixture in \citet{Federrath2010}. The turbulence driving code is available on GitHub \citep{FederrathEtAl2022ascl}. As in \cite{Federrath2021}, the turbulent Mach number is defined as $\mathcal{M} = \sqrt{\langle \vect{u}^2 \rangle /c_s^2}$. Here, the Mach number is $\mathcal{M}=4.1$. The grid consists of $10,\!048^3$ data points stored into $65,\!536$ blocks of 157 $\times$ 314 $\times$ 314 points each. 5~snapshots separated by one eddy turnover time in the fully developed turbulent state were used to average the statistical measurements below.

The surface-scale relation to be described later is obtained from a standard method \citep{Silva2013,Hawkes2012,Thiesset2016,Krug2017}. It consists first in coarse-graining the field variable of interest (here the density) at different filter size $\Delta$. For each filter size, the phase indicator field can then be extracted by thresholding the filtered density field. By doing so, the surface area of the 'filtered' interface can be computed and studied as a function of $\Delta$ to determine the surface-scale relation. 

Here, the filtered density denoted $\overline{\rho}$ is obtained by using a box average over a cubic sub-set of size $\Delta$. We note that the transport equation for the filtered density is the same as the unfiltered one:
\begin{eqnarray}
  \partial_t \overline{\rho}  + \vect{\nabla} \cdot \overline{\rho} \widetilde{\vect{u}} = 0,
\end{eqnarray}
where the Favre-filtered velocity is given by:
\begin{eqnarray}
  \widetilde{\vect{u}} = \frac{\overline{\rho \vect{u}}}{ \overline{\rho}}.
\end{eqnarray}
Therefore, the scale-by-scale budget (Eq.~\eqref{eq:dphi2_eq}) remains formally the same when the Favre-average velocity $\tilde{\vect{u}}$ is used in place of $\vect{u}$. We have investigated different ratios of filter size to the original grid spacing $\Delta_x$, from $\Delta = \Delta_x$ (the unfiltered case) to $32\Delta_x$, resulting in 6~down-sampled datasets composed of $314^3$, $628^3$, $1,\!256^3$, $2,\!512^3$, $5,\!024^3$ and $10,\!048^3$ grid points for $\Delta/\Delta_x=32, 16, 8, 4, 2, 1$. 

The correlation and structure functions are computed using the library \texttt{pyarcher} \citep{Thiesset2020a}. Only spatial separations $\vect{r}$ aligned with the three Cartesian directions ($\vect{e_x}, \vect{e_y}, \vect{e_z}$) were considered and subsequently averaged, taking advantage of isotropy. The separation vector was varying between 1 grid spacing up to half the simulation box size, which is identical to the turbulence driving scale, denoted $L$. The number of sampling pairs is given by $5 ~{\rm (snapshots)} \times 3~{\rm (directions)} \times 10,\!048^3 (\Delta_x/\Delta)^3 ~{\rm (points)}$, and was thus varying between about $10^{8}$ and $10^{13}$ for $\Delta$ between $32 \Delta_x$ and $\Delta_x$, respectively. This was found sufficient to reach statistical convergence of the structure and correlation functions \citep[c.f., respective sampling tests for the structure functions in][]{Federrath2021}.

Seven different values for the density threshold $\rho_{\rm th}$ were chosen: $\rho_{\rm th}/\rho_0 = \{0.1, 0.2, 0.5, 1, 2, 5, 10\}$. In the present simulation, the volume average density $\rho_0 \equiv \langle \rho \rangle = 1$, and hence $\rho_{\rm th}$ will hereafter be given in units of $\rho_0$. The density and the corresponding phase indicator fields for varying $\rho_{\rm th}$ are portrayed in Fig.~\ref{fig:slice}. We note that the density field is highly convoluted with some fluctuations ranging very different scales. When observed at large scales (at the size of the simulation box), dilute regions ($\rho_{\rm th} < 1$) take the form of bulky and agglomerated structures, while dense regions ($\rho_{\rm th} > 1$) are more sparse and filamentary. Neutral density ($\rho_{\rm th} \sim 1$) regions reveal some branched structures. 

\section{Results} \label{sec:results}

\subsection{Volume fraction}

\begin{figure}
  \includegraphics[width=\linewidth]{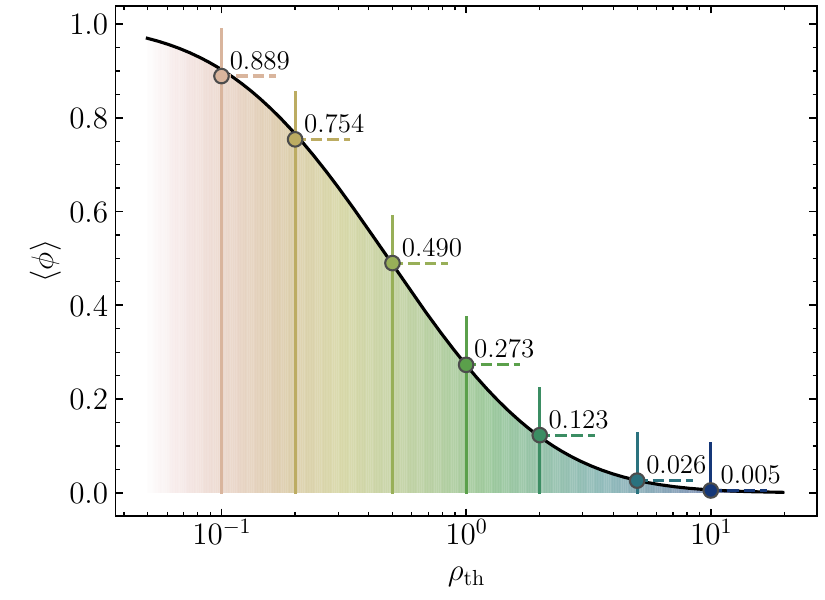}  
  \caption{Volume fraction $\langle \phi \rangle$ for the different values of $\rho_{\rm th}$ (symbols). The black solid line corresponds to the prediction of Eq.~\eqref{eq:phi_rho} with a dispersion parameter $\sigma_s = 1.21$. }\label{fig:phi_rho}
\end{figure}

\cite{Federrath2021} showed that the probability density function of $\rho$ obtained from the present simulation data is well represented by the intermittency model distribution by \citet{Hopkins2013}. However, the value of the intermittency correction was found to be rather small. This suggests that the log-normal distribution provides a good approximation. Assuming that the volume-weighted probability density function of $\rho$ is log-normal, one can easily derive the following expression for the volume fraction $\langle \phi\rangle$:
\begin{eqnarray}
\langle \phi\rangle = \frac{1}{2} \left(1-{\rm erf} \left[\frac{\log \rho_{\rm th} + \frac{1}{2}\sigma_s^2}{\sqrt{2\sigma_s^2}}\right] \right). \label{eq:phi_rho}
\end{eqnarray}
In Eq. \eqref{eq:phi_rho}, '${\rm erf}$' is used to denote the error function and $\sigma_s$ is the volume-weighted dispersion (standard deviation) of $s=\log \rho/\rho_0$. \citet{Federrath2021} measured $\sigma_s=1.21$ for this simulation.

Fig. \ref{fig:phi_rho} gathers the numerical values for $\langle \phi\rangle$ when the iso-density is varied from 0.1 to 10. The colour code for each iso-density value (light for low, dark for high $\rho_{\rm th}$) will be followed in the sequel. The volume fraction is about 90\% for $\rho_{\rm th} = 0.1$ and decreases down to 5\textperthousand~for $\rho_{\rm th} = 10$. The median $\langle \phi \rangle = 0.5$ is obtained for $\rho_{\rm th} = \exp(-\sigma_s^2/2)\approx 0.48$. The prediction assuming a log-normal distribution, Eq. \eqref{eq:phi_rho}, compares favourably well to the numerical data, except maybe at low values of $\rho_{\rm th}$ where the intermittency correction starts to become significant. Fig.~\ref{fig:phi_rho} together with Eq.~\eqref{eq:phi_rho} shows that evaluating $\langle \phi \rangle$ for different $\rho_{\rm th}$ is equivalent to evaluating the (cumulative) probability density function of $\rho$.

\subsection{Structure and correlation functions}

\begin{figure}
  \includegraphics[width=\linewidth]{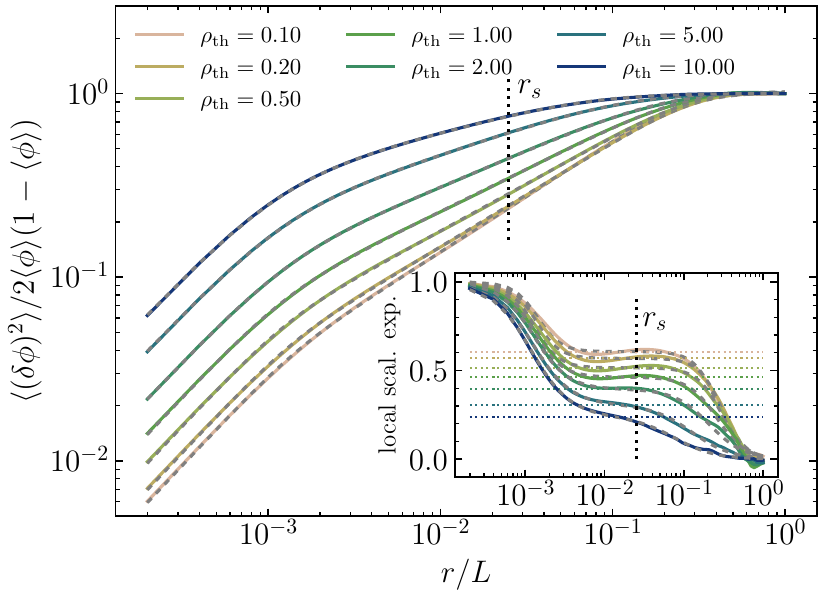} 
  \caption{Scaling of $\langle (\delta \phi)^2 \rangle$ for $\Delta = 1 \Delta_x$, that is $N=10,\!048^3$ and $\rho_{\rm th}$ from 0.1 to 10.0. The inset shows the local scaling exponent $\partial_{\log(r)} \log\langle (\delta \phi)^2 \rangle$. The grey dashed line represents the fit using Eq.~\eqref{eq:dphi2_param}. The sonic scale $r_s$ is shown by the vertical black dotted line. }\label{fig:dphi2_rho} 
\end{figure}

We now analyse the structure function $\langle (\delta \phi)^2\rangle$ for the unfiltered dataset. Results for iso-density values $0.1 \leq \rho_{\rm th} \leq 10$ are presented in Fig.~\ref{fig:dphi2_rho}. The inset represents the local scaling exponent $\partial_{\log(r)} \log \langle (\delta \phi)^2\rangle$. The sonic scale $r_s = 0.025L$ which is the scale at which the local Mach number is equal to one \citep{Federrath2021} is also represented. The structure function is normalised by $2 \langle \phi \rangle(1-\langle \phi \rangle)$, whereas the separation $r$ is normalised by $L$ (the turbulence driving scale). We note that all curves converge to the same plateau when $r \to \infty$ as expected from Eq.~\eqref{eq:dphi2_large_scales}. The local scaling exponent is thus zero in this range of scales. 

Travelling through smaller scales, we note the onset of a power-law behaviour for $\langle (\delta \phi)^2\rangle$. The latter is particularly visible for low $\rho_{\rm th}$. For instance, the local scaling exponent is constant over more than a decade for $\rho_{\rm th} = 0.1$. The power-law range is centred around the sonic scale. We further note that contrary to the velocity structure functions reported by \citet{Federrath2021}, which reveal different scaling exponents below and above the sonic scale, the $\phi$-field structure function exponent is roughly the same in the sub- and supersonic range. The observed power-law behaviour in the intermediate range of scales means that iso-density fields are surface-fractals (c.f., Figures~\ref{fig:fractals} and \ref{fig:Menger_Julia}). The value for the scaling exponent differs depending on the chosen iso-density threshold, which means that the density field cannot be described by a unique surface fractal dimension. Instead, the dimension $D_s$ of iso-density sets increases with $\rho_{\rm th}$. In other words, the fractal content of iso-density surfaces is larger for dense clumps than dilute regions. 

When the separation $r$ tends to smaller values, the structure function $\langle (\delta \phi)^2\rangle$ ceases to follow a power law and the local scaling exponent progressively evolves to reach a value of 1 at very small scales. The scale at which this transition appears is roughly the same irrespective of $\rho_{\rm th}$. This suggests that the inner cutoff, which is the scale below which the iso-surface stops being fractal, is independent of $\rho_{\rm th}$. More details on this aspect will be given later when analysing the surface--size distribution. The fact that $\langle (\delta \phi)^2\rangle$ follows a linear regime when $r \to 0$  means that the iso-density surface is planar when observed at sufficiently small scales. This result is not so intuitive since in supersonic turbulence, the presence of shocks may yield a loss of smoothness of the density field at all scales. It is however unclear if this observed planarity of iso-density surfaces is 'physical' or is due to numerical dissipation.

\citep{Gauding2022} found that $\langle (\delta \phi)^2\rangle$ can be represented by the following parametric expression:
\begin{align}
   \langle (\delta \phi)^2\rangle (r)= \frac{\Sigma {r}}{2} \frac{~~~\left[1 + \left(\frac{{r}}{{\eta_i}}\right)^{\alpha} \right]^{(\xi_s - 1)/\alpha}} {\left[1 + \left(\frac{{r}}{{\eta_o}}\right)^{\alpha}\right]^{\xi_s/\alpha}}. \label{eq:dphi2_param}
\end{align}
This expression accounts for the different regimes described above and makes explicit the dependence of $\langle (\delta \phi)^2\rangle$ to $\Sigma$, $\eta_i$, $\eta_o$ and $\xi_s$. The additional parameter $\alpha$ describes the sharpness of the transition between small, intermediate and large scales. The merit of Eq. \eqref{eq:dphi2_param} is that all relevant features of the phase-indicator function can be inferred unambiguously (without arbitrary adjustments about say the best scaling range) using a least-square fitting. The distributions given by Eq.~\eqref{eq:dphi2_param}, where the parameters $\Sigma$, $\eta_i$, $\eta_o$ and $\xi_s$ are obtained by least-square fitting, are represented by the grey dashed lines in Fig.~\ref{fig:dphi2_rho}. The latter superimpose nearly perfectly on the numerical data, which proves the appropriateness of Eq.~\eqref{eq:dphi2_param} and the least-square method for inferring $\Sigma$, $\eta_i$, $\eta_o$ and $\xi_s$ without any ambiguity. 

\begin{figure}
  \includegraphics[width=\linewidth]{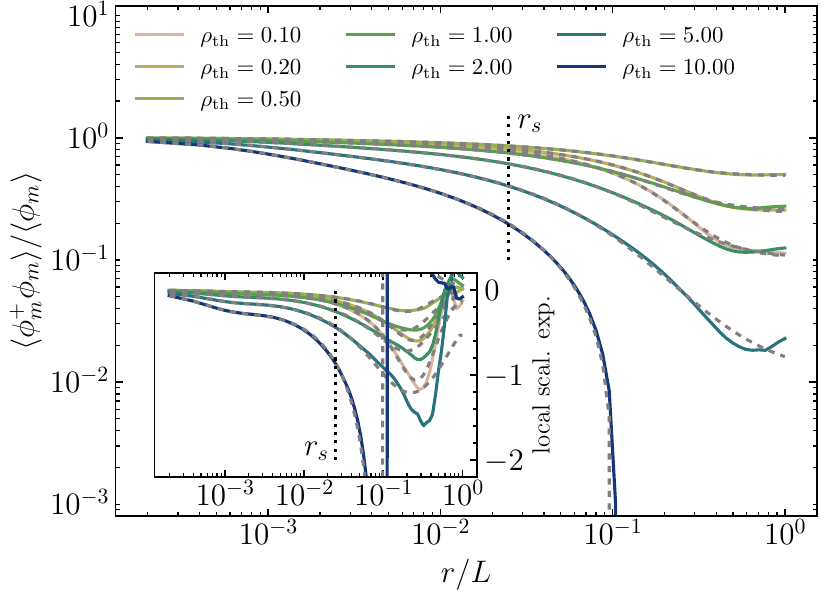} 
  \caption{Scaling of $\langle \phi_m^+ \phi_m \rangle$ for $\Delta = 1 \Delta_x$ ($N=10,\!048^3$) and $\rho_{\rm th}$ from 0.1 to 10.0. The inset shows the local scaling exponent $\partial_{\log(r)} \log\langle \phi_m^+ \phi_m \rangle$. The sonic scale $r_s$ is shown by the vertical black dotted line. The grey dashed line represents the fit using Eq.~\eqref{eq:dphi2_param} and Eq.~\eqref{eq:correl-struct}} \label{fig:corr_rho} 
\end{figure}

In order to assess whether the iso-density field can also be a mass-fractal, we now proceed with the analysis of the correlation function $\langle \phi_m^+ \phi_m\rangle$ for varying $\rho_{\rm th}$. We have chosen to compute and show the results for the correlation function of the minority phase $\phi_m$ simply because it is the one that has the best prospect of showing a power-law behaviour. Indeed, since the correlation varies between $\langle \phi \rangle$ and $\langle \phi \rangle^2$ at asymptotically small and large scales, respectively, it is expected that the scaling range is maximised when $\langle \phi \rangle$ is the smallest, hence for the minority phase. Despite this caution, the computed correlation functions presented in Fig.~\ref{fig:corr_rho} do not reveal any range of scales complying with a power-law relation. This is also confirmed by the local scaling exponent $\partial_{\log(r)} \log \langle \phi_m^+ \phi_m\rangle$, which is displayed in the inset. There is only a hint of a plateau at scales much smaller than $r_s$ for dense regions (for $\rho_{\rm th} = 10$) but the value obtained for $D_m$ is about 2.8, which is quite close to 3 (the value obtained for a pure surface-fractal). The conclusion here is that iso-density fields of supersonic isothermal turbulence are not mass-fractals, except maybe for very dense clumps. This confirms the direct visualisation provided in Fig.~\ref{fig:dphi2_asymptotics}, which reveals that at intermediate scales, the iso-density field is clearly more surface-fractal than mass-fractal.

Given that the set under consideration here is a surface-fractal, it is expected that the mass--size relation is $M(r) \sim r^3$. This appears in disagreement with the consensus based on robust numerical \citep{Federrath2009,Kritsuk2007,Audit2010} and observational evidence in molecular clouds \citep[see for instance the review by][and references therein]{RomanDuvalEtAl2010,Hennebelle2012} for a mass fractal dimension $D_m<3$. The origins for this disagreement are not yet clear. One first explanation is that the present method based on the correlation of the phase indicator field does not measure the same dimension as the one inferred by the methods of \citet{Kritsuk2007} and \cite{Federrath2009}, which consists of measuring the mass contained in boxes of size $r$ centred around the density peaks. Second, \citet{Kritsuk2007} and \citet{Federrath2009} use the original density field, while here we consider a thresholded version thereof. In other words, in the present framework, the mass can be viewed as being composed of a sum of weights, which are equal to either 0 or 1, while in \citet{Kritsuk2007} and \citet{Federrath2009}, the mass is obtained after spatial integration of $\rho$ over a box of size $r$. In \citet{Audit2010} the mass--size relationship is also inferred from a thresholded density field, but the mass is computed for each individual connected object and the scale is defined from the largest eigenvalue of the inertia tensor defined for each structure. These differences in the definition of both the mass and the scale are likely to explain the difference between our conclusion and the one reported by \citet{Audit2010}.

\subsection{Scale-space flux}

\begin{figure}
  \centering
  \includegraphics[width=\linewidth]{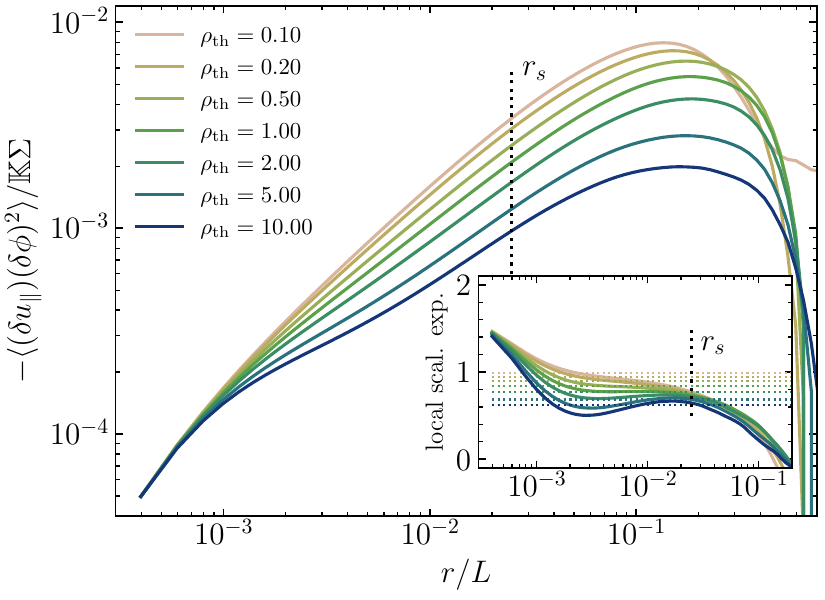} 
  \caption{Scaling of $\langle (\delta u_\parallel) (\delta \phi)^2 \rangle$ for $\rho_{\rm th}$ from 0.1 to 10.0. The inset shows the local scaling exponent. The horizontal dotted lines are the predicted scaling exponents $\xi_u + \xi_s$. } \label{fig:dudphi2_rho} 
\end{figure}

The scale distribution of the flux $\langle (\delta u_\parallel) (\delta \phi)^2\rangle$ for different values of the density threshold is shown in Fig.~\ref{fig:dudphi2_rho}. We have chosen to normalise $\langle (\delta u_\parallel) (\delta \phi)^2\rangle$ by $\mathbb{K}\Sigma$. Using this normalisation, all curves collapse at small scales, which is in agreement with Eq.~\eqref{eq:dudphi2_small_scales}.

As a first remark, we note on Fig.~\ref{fig:dudphi2_rho} that $\langle (\delta u_\parallel) (\delta \phi)^2\rangle$ is negative irrespective of the probed scale $r$. This means that the quantity $\langle (\delta \phi)^2\rangle$ is transported towards smaller scales, following a direct cascade process. Hence, our results are consistent with the classical view that turbulence acts in stirring, stretching and folding the density field, thereby increasing its structural content. Broadly speaking, the turbulent velocity field acts in concentrating more interface into smaller volumes. The budget given by Eq.~\eqref{eq:dphi2_eq} then suggests that the velocity divergence counteracts this effect and is responsible for the expansion of density structures. 

Second, we observe that although scaled in terms of strain rate, which for iso-density fields plays the same role as the scalar or the kinetic energy dissipation rate for the scalar variance or kinetic energy \citep{Thiesset2020,Gauding2022}, the flux depends quite significantly on the density threshold. This suggests that the dense, neutral and dilute regions do not equally interact with the turbulent velocity field. By contrast, the flux decreases with increasing $\rho_{\rm th}$.

Thirdly, at intermediate scales, we observe the onset of a power-law behaviour for the flux term. This is better illustrated by the evolution of the local scaling exponent, shown in the inset of Fig.~\ref{fig:dudphi2_rho}. A careful examination reveals that the scaling range is narrower than the one observed for $\langle (\delta \phi)^2\rangle$ and appears at scales smaller than the sonic scale, which corresponds to the subsonic part of the turbulent spectrum. In this region of scales, \cite{Federrath2021} found that $\langle(\delta u_\parallel)^2\rangle^{1/2}\sim r^{\xi_u}$ where $\xi_u = 0.39$ \footnote{To be precise, the value of 0.39 for the exponent reported by \cite{Federrath2021} was obtained by summing longitudinal and transverse velocity fluctuations. We have checked that it was the same for the longitudinal component only.}. As per \citet{Gauding2022}, assuming that the flux scales as:
\begin{eqnarray}
  \langle (\delta u_\parallel) (\delta \phi)^2\rangle \sim \langle(\delta u_\parallel)^2\rangle^{1/2} \langle(\delta \phi)^2\rangle, 
\end{eqnarray}
we find that $\langle (\delta u_\parallel) (\delta \phi)^2\rangle$ should scale as $r^{\xi_u+\xi_s}$. This prediction is plotted as the horizontal dotted lines in the inset of Fig.~\ref{fig:dudphi2_rho}. Although not perfect, the agreement between this rather crude reasoning and numerical data is satisfactory. In particular, it reproduces rather well the decreasing evolution of the exponent for increasing $\rho_{\rm th}$. 

Finally, we note that similarly to the observations of \cite{Gauding2022} for iso-scalar surfaces in incompressible turbulence, there is not or only a very limited range of scales where the transfer rate is constant, that is where $\langle (\delta u_\parallel) (\delta \phi)^2\rangle \sim r$. The same observation was carried out by \cite{Ferrand2020} for the scale-by-scale transfer rate of kinetic energy. According to \cite{Ferrand2020}, the non-constant energy transfer can be attributed to either shocks/discontinuities, yielding a loss of smoothness of the velocity field \citep{Duchon2000,Saw2016,Galtier2018,Dubrulle2019} or to non-local effects of the large-scale numerical forcing. We may also conjecture that the present numerical resolution, although unprecedented, is still not sufficient to observe a constant transfer of transported quantities (density, iso-density, or velocity) in the inertial range. This question is left unanswered.

\subsection{Filtered quantities}

We now proceed with the two-point statistical analysis of the phase indicator $\phi$ defined from filtered density $\overline{\rho}$ at varying filter size $\Delta$ and thresholded at a given $\rho_{\rm th}$. For the sake of conciseness, we do not consider all above values for $\rho_{\rm th}$, but focus on $\rho_{\rm th} = 1$ only. Qualitatively similar observations were carried out for the other iso-density values. 

\begin{figure}
  \includegraphics[width=\linewidth]{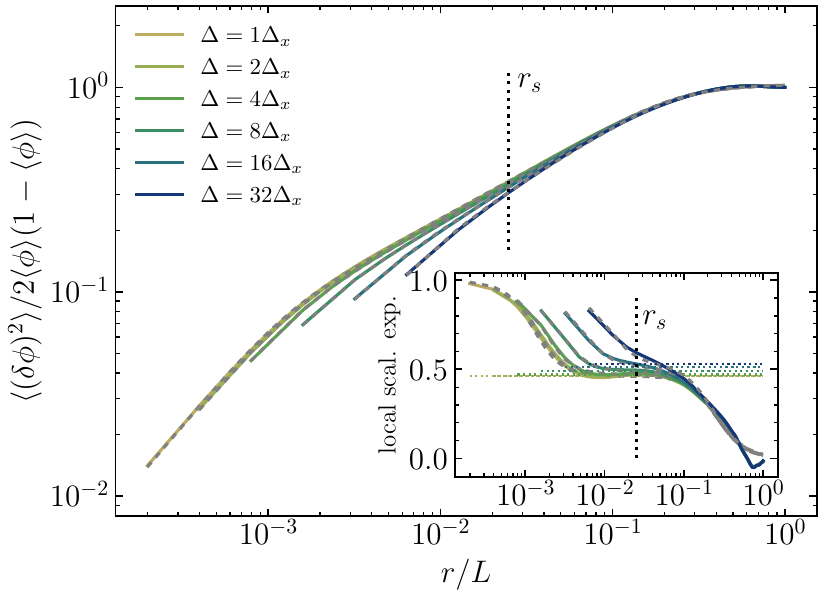} 
  \caption{Same as Fig.~\ref{fig:dphi2_rho}, but for $\rho_{\rm th}=1.0$ and $\Delta$ varying from 1 to $32\Delta_x$.} \label{fig:dphi2_npt}
\end{figure}

The second-order structure function $\langle (\delta \phi)^2\rangle$ for $\Delta/\Delta_x$ varying from 1 to 32 is plotted in Fig. \ref{fig:dphi2_npt}. As expected, increasing the filter size results in an earlier cutoff of $\langle (\delta \phi)^2\rangle$ at small scales. The direct consequence is that the surface density $\Sigma$ decreases with increasing $\Delta$. More insights into this behaviour will be given in the next section, which focuses on the surface--size relation. 

We note that the differences between the distributions for different $\Delta$ are rather small for $\Delta < 4 \Delta_x$, but becomes significant for $\Delta \geq 8 \Delta_x$. However, the behaviour of $\langle (\delta \phi)^2\rangle$ at large scales remains unaffected by the filter size, which indicates that \textit{(i)} the measured volume fraction $\langle \phi\rangle$, and \textit{(ii)} the outer cutoff $\eta_o$ do not depend on $\Delta$. This may remain valid as long as $\Delta \ll \eta_o$. In Fig.~\ref{fig:dphi2_npt}, the fitted parametric expression given by Eq.~\eqref{eq:dphi2_param} is plotted as the grey dashed lines. Here again, it shows that the proposed expression for $\langle (\delta \phi)^2\rangle$ agrees very well with the numerical data. It can thus be used with confidence to unambiguously infer the geometric quantities $\Sigma$, $\eta_o$, $\eta_i$ and $\xi_s$, even when the filter size is large, resulting in a restricted scaling range. 

More insights can be provided by looking at the local scaling exponent of $\langle (\delta \phi)^2\rangle$, which is shown in the inset of Fig.~\ref{fig:dphi2_npt}. It reveals that the extent of the power-law range becomes narrower with increasing $\Delta$. The local scaling appears to stabilise around a plateau at intermediate scales whose value depends on $\Delta$. For the present $\rho_{\rm th}$, the local scaling exponent obtained by fitting the numerical data using the parametric expression, Eq.~\eqref{eq:dphi2_param}, increases from about 0.46 for $\Delta = \Delta_x$ to 0.52 for $\Delta = 32\Delta_x$. Therefore, these variations although measurable, are much smaller than the variations associated with different values of $\rho_{\rm th}$ (c.f.~Fig.~\ref{fig:dphi2_rho}). As a first approximation, one can thus assume that $\xi_s$ depends only on the chosen iso-density, but remains constant, independently of the resolution $\Delta$. Speculatively, the variation of $\xi_s$ with $\Delta$, though small, is evidence for a multi-fractal density field characterised by a fractal dimension that depends on the probed scale \citep{Chappell2001}. Here, it appears that when observed with a finer resolution, the fractal dimension is increasing, meaning that the interface is more tortuous, more space-filling. This statement about the multi-fractal character of iso-density sets, is rather hasty at this stage and a deeper analysis is required for being confirmed. This is left for future investigations.

\begin{figure}
  \centering
  \includegraphics[width=\linewidth]{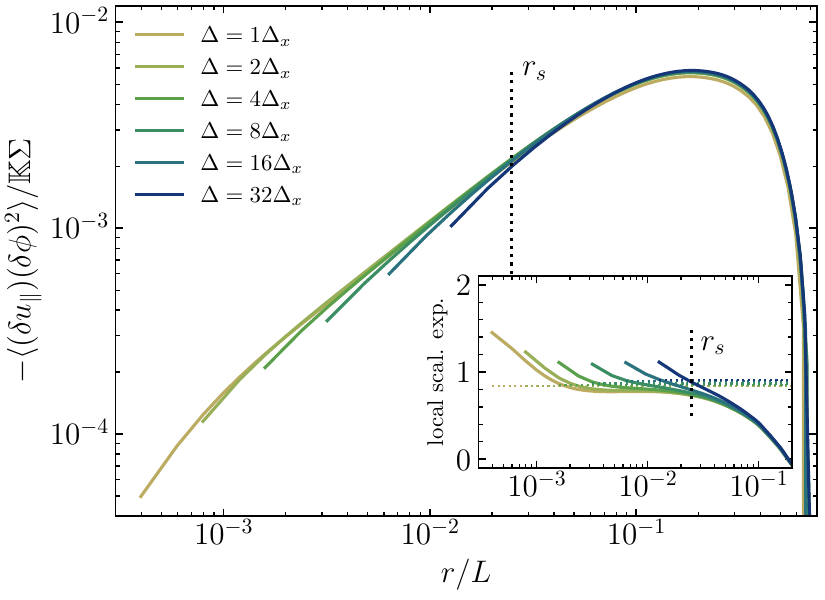} 
  \caption{Same as Fig. \ref{fig:dudphi2_rho}, but for $\rho_{\rm th}=1.0$ and $\Delta$ varying from 1 to $32\Delta_x$.} \label{fig:dudphi2_npt}
\end{figure}

The flux $\langle (\delta u_\parallel) (\delta \phi)^2\rangle$ for varying filter size is presented in Fig.~\ref{fig:dudphi2_npt}. Here again, we plot the results only for $\rho_{\rm th} = 1.0$, but qualitatively similar conclusions were drawn for the other iso-density values. We again normalise the flux by $\mathbb{K}\Sigma$, this quantity being estimated for $\Delta = 1\Delta_x$. Increasing the filter size $\Delta$ results again in a faster drop of the flux in the small-scale limit. This means that the measured strain rate $\mathbb{K}\Sigma$ decreases with increasing filter size. Comparing Figs.~\ref{fig:dudphi2_npt} and \ref{fig:dphi2_npt}, we also note that the effect of $\Delta$ is somewhat more visible than it was for $\langle (\delta \phi)^2\rangle$ and becomes substantial for $\Delta \geq 2 \Delta_x$. However, the distributions at large scales remain unchanged. The local scaling exponent for the flux $\langle (\delta u_\parallel) (\delta \phi)^2\rangle$ (see the inset in Fig.~\ref{fig:dudphi2_npt}) stabilises around a plateau whose value is only weakly affected by the filter size. As a result of a larger inner cut-off $\eta_i$, the extent of the scaling range diminishes measurably when $\Delta$ is increased. 

In short, as long as $\eta_i < \Delta \ll \eta_o$, the large-scale distributions will not be affected by the filtering operation. In the opposite limit, that is for $\Delta \ll \eta_i $, both $\langle(\delta \phi)^2\rangle$ and $\langle (\delta u_\parallel) (\delta \phi)^2\rangle$ are independent of $\Delta$ over the entire range of scales. For $\eta_i < \Delta < \eta_o$, the measured inner cutoff increases with $\Delta$ (more insights into this evolution is given in the next section), while the measured fractal dimension is only weakly affected by the filter size. 

\subsection{Surface-scale relation}

For surface-fractals, the surface density $\Sigma$ is expected to follow the relation \citep{Sreenivasan1989}:
\begin{eqnarray}
  \Sigma = \kappa_f \left(\frac{\eta_o }{\eta_i}\right)^{D_s-2}, \label{eq:sigma}
\end{eqnarray}
where $\kappa_f$ is the fractal pre-factor, which has here the dimension of $\Sigma$ (units of inverse length). It depends only on $\rho_{\rm th}$ but not on $\Delta$. The quantities involved in Eq.~\eqref{eq:sigma} have been quantified by fitting the numerical estimations of $\langle (\delta \phi)^2\rangle$ with the parametric expression of Eq.~\eqref{eq:dphi2_param}.

\begin{figure}
  \includegraphics[width=\linewidth]{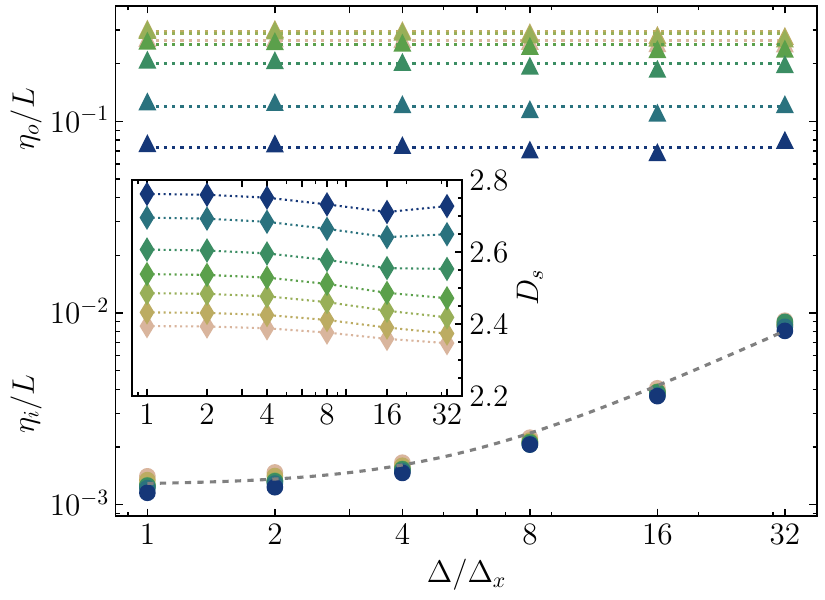}  
  \caption{Inner cutoff $\eta_i$ (circles) and outer cutoff $\eta_o$ (triangles), together with the fractal dimension $D_s$ (diamonds) as a function of $\Delta/\Delta_x$ for different $\rho_{\rm th}$ ranging from 0.1 to 10.} \label{fig:etai_etao}
\end{figure}

Let us first focus on the dependence of $\eta_o$ with respect to $\Delta$ for different $\rho_{\rm th}$. The latter is illustrated by triangles in Fig.~\ref{fig:etai_etao}. The limit of Eq.~\eqref{eq:dphi2_param} when $r \to \infty$, together with Eq.~\eqref{eq:dphi2_large_scales}, reveals that
\begin{eqnarray}
  2 \langle \phi \rangle (1-\langle \phi \rangle) = \frac{\Sigma}{2} \left(\frac{\eta_o}{\eta_i} \right)^{2-D_s} \eta_o.
\end{eqnarray}
By virtue of Eq.~\eqref{eq:sigma}, we then have
\begin{eqnarray}
  \eta_o = 4 \kappa_f^{-1} \langle \phi \rangle (1-\langle \phi \rangle). \label{eq:kappa_f}
\end{eqnarray}
Thus, there exists a close link between the outer cutoff, the volume fraction and the fractal pre-factor. Since the measured volume fraction $\langle \phi \rangle$ does not depend on $\Delta$ (see also Fig.~\ref{fig:dphi2_npt}), and that by definition, $\kappa_f$ is function of $\rho_{\rm th}$ only, we thus predict that the outer cutoff $\eta_o$ should be constant with respect to $\Delta$. This prediction is confirmed by Fig.~\ref{fig:etai_etao}. 

The fractal dimension $D_s = 3-\xi_s$ is displayed in the inset of Fig.~\ref{fig:etai_etao} for the different filter sizes and different density iso-values investigated here. We note that, as observed previously for $\rho_{\rm th} = 1.0$, the fractal dimension depends much more on $\rho_{\rm th}$ than on the filter size. Although small in amplitude, there seems to be increasing evolution of $D_s$ when $\Delta$ decreases, before reaching an approximately constant value for $\Delta \lesssim 4 \Delta_x$. Overall, assuming that $D_s$ is a function of $\rho_{\rm th}$ only, seems to be a reasonable assumption.

The evolution of the inner cutoff $\eta_i$ for increasing filter size is also shown in Fig.~\ref{fig:etai_etao}. The first immediate conclusion is that $\eta_i$ depends on $\Delta$, but remains the same in dilute, dense, or neutral regions. Interestingly, \citet{Gauding2022} also found that $\eta_i$ of passive scalar iso-surfaces in incompressible turbulence are independent of the iso-scalar value. These authors found that $\eta_i$ can be predicted as the scale at which there is equilibrium between production (associated with the turbulent strain rate) and destruction of surface curvature (in their case related to scalar diffusion). The same reasoning applied to the present flow configuration suggests that the equilibrium between production and destruction of interface, associated here with turbulent straining and velocity divergence, respectively, is reached at the same scale irrespective of the probed iso-density value.  

The second observation is that for small values of $\Delta$, $\eta_i$ is constant and starts increasing when $\Delta \gtrsim 4 \Delta_x$. For large values of $\Delta$, it is reasonable to assume that $\eta_i \sim \Delta$, while for any resolution $\Delta \ll \eta_i$, the interface tortuousness is fully resolved and the measured $\eta_i$ is constant and is equal to $\eta_i^{u}$: the unfiltered inner cutoff. These two distinct behaviours:
\begin{eqnarray}
  \eta_i(\Delta)  \begin{cases}
              = \eta_i^u ~~ {\rm if} ~\Delta \ll \eta_i^u  \\
              \sim \Delta ~~ {\rm if ~} \Delta \gg \eta_i^u, 
           \end{cases}
\end{eqnarray}
can be combined into a single expression of the form
\begin{eqnarray}
  \eta_i(\Delta)  = \eta_i^u \left[1+\left(\frac{a\Delta}{\eta_i^u}\right)^b\right]^{1/b}, \label{eq:eta_i}
\end{eqnarray}
where $a$ and $b$ are two constants of order unity. In Fig.~\ref{fig:etai_etao}, Eq.~\eqref{eq:eta_i} is compared to the numerical data. We find that $a = 1.25$ and $b=2$ provide satisfactory results, although we did not seek for the most suitable values (we did not perform any least-square fit). The unfiltered inner cutoff $\eta_i^u$ is found to be $\eta_i^u = (1.26  \pm 0.04) \times 10^{-3} L \approx 6\Delta_x$. It is worth stressing that here, the inner cutoff is set artificially by numerical dissipation. In reality, the inner cutoff scale may be of the same order as the viscous scales \citep{Gauding2022}. 

\begin{figure}
  \includegraphics[width=\linewidth]{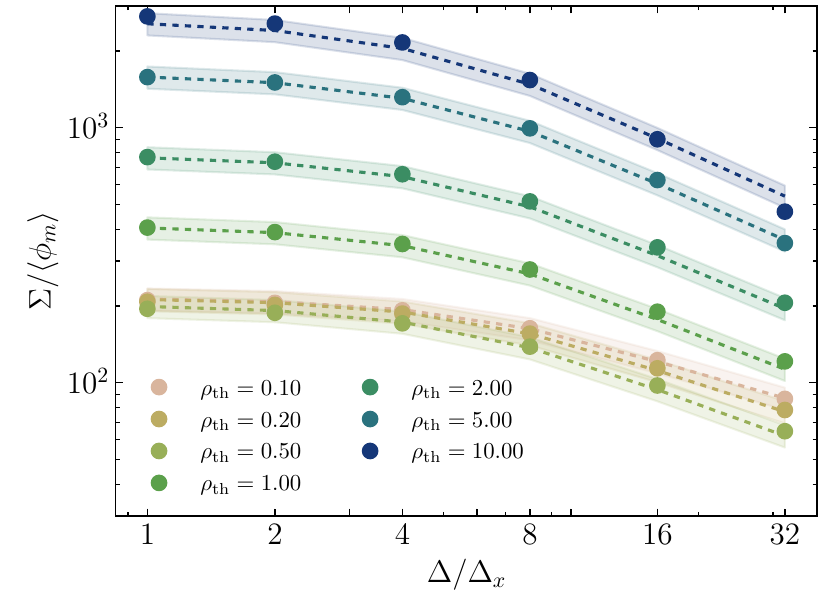}  
  \caption{Specific surface $\Sigma/\langle\phi_m\rangle$ as a function of $\Delta/\Delta_x$ for $\rho_{\rm th}$ ranging from 0.1 to 10. The dashed lines show the prediction with $\Sigma = \kappa_f (\eta_o/\eta_i)^{D_s-2}$. The coloured filled regions correspond to a relative error of $\pm 10\%$.} \label{fig:sigma}
\end{figure}

In Fig.~\ref{fig:sigma}, we plot the specific surface density $\Sigma_s$, defined as the ratio between the surface area of the interface and the volume of the minority phase. The specific surface density is thus related to the above defined surface density $\Sigma$ by $\Sigma_s = \Sigma/\langle \phi_m\rangle = \Sigma / \min(\langle \phi\rangle , 1-\langle \phi\rangle)$. The values for $\Sigma_s$ are computed from the fitted parametric expression, Eq.~\eqref{eq:dphi2_param}, and are represented with the circle symbols in Fig.~\ref{fig:sigma}. A first observation is that the specific surface for the dense regions is much higher than that of the neutral or dilute regions. In other words, dense regions are the ones for which the ratio between the surface area and the volume is the highest. This is consistent with our previous conclusions that the surface fractal dimension $D_s$ is larger for high density regions, meaning that the interface is more corrugated. 

The evolution of $\Sigma_s$ with respect to the filter size is qualitatively similar irrespective of the iso-density. For small values of the filter size, for $\Delta < \eta_i^u$, it reaches a constant value and decreases for any filter size $\Delta > \eta_i^u$. For large values of $\Delta$, we observe the onset of a power law for the specific surface density with respect to $\Delta$. The evolution of the filtered specific surface density can be well predicted by the surface-fractal model. Indeed, using Eq.~\eqref{eq:kappa_f} to express the fractal pre-factor $\kappa_f$ in terms of $\langle \phi \rangle$ and $\eta_o$, we end up with the following expression for $\Sigma_s$:
\begin{eqnarray}
  \Sigma_s = 4 \frac{\langle \phi \rangle (1-\langle \phi \rangle)}{\eta_o \langle \phi_m \rangle} \left(\frac{\eta_o }{\eta_i}\right)^{D_s-2}. \label{eq:sigma_model}
\end{eqnarray}
This expression is tested in Fig.~\ref{fig:sigma}. Predictions are illustrated using the dashed lines, where $\eta_o$, $\eta_i$ and $D_s$ were also extracted from the fitting procedure. We note a close agreement irrespective of the iso-density value. Some discrepancies between the model and the numerical data are in the range $\pm 10\%$, which is illustrated by the coloured filled regions. The adequacy of Eq.~\eqref{eq:sigma_model} is further evidence that the iso-density sets are surface-fractals. 

The power-law behaviour for $\Sigma_s$ at large $\Delta$ can be derived. For this purpose, one needs \textit{(i)} to recall that $\eta_o$ and $\langle \phi \rangle$ depend only on $\rho_{\rm th}$, while \textit{(ii)} $\eta_i$ is proportional to $\Delta$ in the limit of large $\Delta$, and finally \textit{(iii)} assume that $D_s$ is constant with respect to $\Delta$. With this, we obtain that $\Sigma_s \sim \Delta^{2-D_s}$, which is the known surface--size relation for surface fractals. This simple expression (and the more detailed one given by Eq.~\eqref{eq:sigma_model}) can be readily used to compare the iso-density surface area estimated using different numerical and/or observational resolutions at the condition that $D_s$ is known. Eq.~\eqref{eq:sigma_model} also requires the parameter $\eta_i^u$ to be known. Conversely, this relation can also be used to estimate the surface fractal dimension and the inner cutoff, using numerical and/or observational data filtered at different resolutions.

In case of observational data, since we dispose only of integrated visualisations, a hypothesis is required for the third dimension. One could for example assume fractal isotropy and then look at the perimeter of the iso-line formed by a given iso-value \citep[as was done for instance by][]{Federrath2009}, replacing the exponent by $1-D_s$ in Eq.~\eqref{eq:sigma_model} \citep[see also work by][]{SanchezEtAl2005,BeattieFederrathKlessen2019,BeattieEtAl2019b}. Therefore, this fractal surface-size relation could help in providing insights into the structural content of the density field inferred from either numerical simulations or observations.
This question could be addressed in a follow-up study.

\subsection{Relation to the virial theorem}

Following \citet{Ballesteros-Paredes1999} and \citet{Dib2007}, let the virial theorem be applied to a volume $V_\rho$ enclosed by the iso-surface $\rho(\vect{x},t)=\rho_{\rm th}$. The volume boundary is denoted $\partial V_\rho$. The Lagrangian formulation for the virial theorem can be written in symbolic form \citep{Chandrasekhar1953,McKee1992,Ballesteros-Paredes2006} as
\begin{eqnarray}
\frac{1}{2}\ddot{I}_L  = 2(\mathcal{E}_{\rm kin} + \mathcal{E}_{\rm int} ) + \mathcal{E}_{\rm mag} - 2 \mathcal{T}_{\rm int} - \mathcal{T}_{\rm mag} - \mathcal{W}, \label{eq:virial}
\end{eqnarray}
where $I_L$ is the moment of inertia of the volume under consideration. The terms denoted with the letter $\mathcal{E}$ in Eq.~\eqref{eq:virial} are the volume integrals over the kinetic, internal and magnetic energy density, respectively, given by
\begin{subequations}
\begin{eqnarray}
  &&\mathcal{E}_{\rm kin} = \frac{1}{2} \int_{V_\rho} \rho u^2 {\rm d}V, \\
  &&\mathcal{E}_{\rm int} = \frac{3}{2} \int_{V_\rho} p {\rm d}V, \\
  &&\mathcal{E}_{\rm mag} = \frac{1}{8\pi} \int_{V_\rho} B^2 {\rm d}V.
\end{eqnarray}
\end{subequations}
The terms denoted with the letter $\mathcal{T}$ are surface integrals:
\begin{subequations}
  \begin{eqnarray}
    &&\mathcal{T}_{\rm int} = \frac{1}{2} \int_{\partial V_\rho} p \vect{x} \cdot \vect{n}~ {\rm d}S, \\
    &&\mathcal{T}_{\rm mag} = \frac{1}{4\pi} \int_{\partial V_\rho} \mathcal{B}~{\rm d}S,
  \end{eqnarray}
\end{subequations}
and represents the surface integrated pressure and magnetic stresses, respectively. The quantity $\mathcal{B}$ is given by:
\begin{eqnarray}
  \mathcal{B} = \vect{x}\cdot \left(\vect{B}\vect{B}- \frac{1}{2}B^2 \vect{I} \right) \cdot \vect{n}.
\end{eqnarray}
The last term in Eq.~\eqref{eq:virial} represents the effect of gravity and reads:
\begin{eqnarray}
  \mathcal{W}_{\rm int} =  \int_{V_\rho} \vect{x} \cdot \vect{\nabla} \Psi  {\rm d}V,
\end{eqnarray}
where $\Psi$ is the gravitational potential. In previous equations, positions and velocity are defined relative to the positions and velocity of the centre of mass. Dividing all terms in Eq.~\eqref{eq:virial} by $V_\rho = \langle \phi \rangle V$, the three volume integrals in Eq.~\eqref{eq:virial} can be rewritten as:
\begin{eqnarray}
  && \frac{1}{\langle \phi \rangle V} (2\mathcal{E}_{\rm kin} + 2\mathcal{E}_{\rm int} + \mathcal{E}_{\rm mag} - \mathcal{W})  =  \nonumber \\
  && ~~~~~\langle \rho u^2 \rangle_\rho + 3 \langle p \rangle_\rho + \frac{1}{8 \pi} \langle B^2 \rangle_\rho + \langle \vect{x} \cdot \vect{\nabla} \Psi \rangle_\rho,
\end{eqnarray}
where the brackets $\langle \bullet \rangle_\rho $ stand for a volume average over $V_\rho$:
\begin{eqnarray}
   \langle \bullet \rangle_\rho = \frac{1}{V_\rho} \int \bullet {\rm d}V = \frac{1}{\langle \phi \rangle V} \int \bullet {\rm d}V.
\end{eqnarray}
Similarly, it is convenient to use the surface area weighted average:
\begin{eqnarray}
  \langle \bullet \rangle_s = \frac{1}{S} \int_{\partial V_\rho} \bullet ~{\rm d}S = \frac{1}{\Sigma V} \int_{\partial V_\rho} \bullet ~{\rm d}S,
\end{eqnarray}
which allows the surface integrals to be rewritten as
\begin{eqnarray}
\frac{1}{\langle \phi \rangle V} (2\mathcal{T}_{\rm int} + \mathcal{T}_{\rm mag}) = \frac{1}{\langle \phi \rangle} \left[ \langle p \vect{x} \cdot \vect{n}\rangle_s + \frac{1}{4\pi}\langle \mathcal{B} \rangle_s \right] \Sigma.
\end{eqnarray}
Finally, the virial theorem can be recast in the form
\begin{eqnarray}
\frac{1}{2 V}\ddot{I}_L  &=& 
\left[\langle \rho u^2 \rangle_\rho + 3 \langle p \rangle_\rho + \frac{1}{8 \pi} \langle B^2 \rangle_\rho + \langle \vect{x} \cdot \vect{\nabla} \Psi \rangle_\rho\right] \langle \phi \rangle  \nonumber \\
&-& \left[ \left\langle p \vect{x} \cdot \vect{n}\right\rangle_s + \frac{1}{4\pi}\left\langle \mathcal{B} \right\rangle_s\right] \Sigma. \label{eq:virial_phi_sigma}
\end{eqnarray}
As before, $V_\rho/V = \langle \phi \rangle$ is the volume occupied by the volume enclosed by the surface $\rho(\vect{x},t)=\rho_{\rm th}$ divided by $V$ (the observed volume), and $\Sigma = S/V$ is the surface density of this surface. A similar expression for the virial theorem in the Eulerian form \citep{McKee1992} can de derived. 

The merit of formulating the virial theorem in the form proposed here is that it draws connections between the geometry of the density field ($\langle \phi \rangle$ and $\Sigma$) and its dynamics (the terms within the brackets). It further allows the gravitational equilibrium to be probed for each iso-density value separately. The different terms in Eq.~\eqref{eq:virial_phi_sigma} cannot be inferred from observations without invoking some simplifying assumptions. They can, however, be estimated from numerical simulations \citep{Ballesteros-Paredes1999,Dib2007}, thereby opening interesting perspectives to explore the relation between the geometry and the dynamics of the ISM. 

\section{Conclusion} \label{sec:conclusion}

The present work aims at exploring the role of supersonic turbulence in shaping the microstructure of the density field. For this purpose, we propose using a two-point statistical analysis of the phase indicator field $\phi$ defined by iso-density sets, Eq.~(\ref{eq:phi}). The asymptotic behaviour for the correlation and structure functions of iso-sets, at small, intermediate and large-scales, are derived theoretically and discussed. These relations revealed that the two-point statistics of $\phi$ depend on some geometric features such as the volume-fraction, the surface density, the curvature, and the fractal characteristics of the gas density field. It is also shown that comparing the correlation and structure function at intermediate scales, allows one to assess whether the medium under consideration is a mass-fractal or a surface-fractal, with important consequences for the mass--size relation. We also derive the transport equation for the correlation and structure functions, emphasising the role of velocity and velocity dilatation in the structural evolution of the density field.

This framework is here appraised using data from highly resolved numerical simulations of supersonic isothermal turbulence. We consider both the original dataset together with the associated filtered quantities in order to establish the surface--size relation of the iso-density fields. 

Our results indicate that iso-density sets of supersonic isothermal turbulence are surface-fractals rather than mass-fractals, except maybe for the very dense regions. The surface-fractal dimension $D_s$ depends significantly on the iso-density value, and increases with increasing density threshold $\rho_{\rm th}$. The consequence is that the specific surface density is higher in the dense regions compared to the dilute or neutral regions. The surface-fractal dimension varies only slightly with the filter size. As a first approximation, it is thus reasonable to assume that $D_s$ depends only on $\rho_{\rm th}$. A direct consequence of the surface fractality of iso-density fields is a model to predict the surface density as a function of the resolution scale. This model could be used to assess the surface-fractal dimension $D_s$ from observations and numerical simulations of the interstellar medium.

The transport equation for the correlation and structure functions reveals that the turbulent cascade and dilatation are two competing effects. The numerical simulation data indicate that the flux in the cascade is negative, meaning that the transfer is occurring from large to small scales (direct cascade). In other words, irrespectively of the probed scale, turbulence acts in concentrating more interface into smaller volumes. Here, dilatation compensates turbulent straining, such that a steady state can be reached. A local scaling range is observed for the flux of iso-density in scale-space, with an exponent that appears to depend on both the velocity and the iso-density power-law scaling. In agreement with \cite{Ferrand2020} for the cascade of kinetic energy, we do not find a clear range of scales complying with a constant scale-transfer (linear flux). As anticipated by \cite{Ferrand2020}, the loss of smoothness of the velocity field and non-local effects of the velocity forcing could explain this observation. We may also conjecture that, though already fine, the resolution is still not fine enough to observe a clear separation of scales between the forcing at large scales and numerical dissipation at small scales, for the phase indicator field $\phi$. 

Finally, a formulation for the virial theorem in terms of $\phi$ is developed, which makes explicit the relation between the geometry of the density field (the volume and surface density) and its dynamics. This new formulation together with the proposed framework based on the two-point statistics of the phase indicator may offer interesting perspectives to better understand the dynamics of the ISM. 

\vspace{16pt}
FT acknowledges A.~Poux and M.~Gauding for their help in the development of the post-processing routines. We thank J.~Yon from the CORIA laboratory for providing the illustration of a soot particle shown in Fig.~\ref{fig:fractals}. CF acknowledges funding provided by the Australian Research Council (Future Fellowship FT180100495 and Discovery Projects~DP230102280), and the Australia-Germany Joint Research Cooperation Scheme (UA-DAAD). We acknowledge computational time granted by the CRIANN (project 2018002), by the J\"ulich Supercomputing Centre (project instahype). We further acknowledge high-performance computing resources provided by the Australian National Computational Infrastructure (grant ek9) and the Pawsey Supercomputing Centre (project~pawsey0810) in the framework of the National Computational Merit Allocation Scheme and the ANU Merit Allocation Scheme, and by the Leibniz Rechenzentrum and the Gauss Centre for Supercomputing (grants pr32lo, pr48pi, pn73fi, and GCS Large-scale projects~10391 and 22542). The simulation software FLASH was in part developed by the DOE-supported Flash Center for Computational Science at the University of Chicago.

\appendix
\section{Simulations at different resolutions}

In addition to analysing data at different filter size $\Delta_x$, we computed the same quantities as in Figs. \ref{fig:etai_etao} and \ref{fig:sigma} using different simulations at different resolutions $\Delta_x$. These simulations were performed using different grid size of $10048^3$, $5024^3$, $2512^3$, $1256^3$ and $628^3$ grid points, respectively. 

\begin{figure}
  \includegraphics[width=\linewidth]{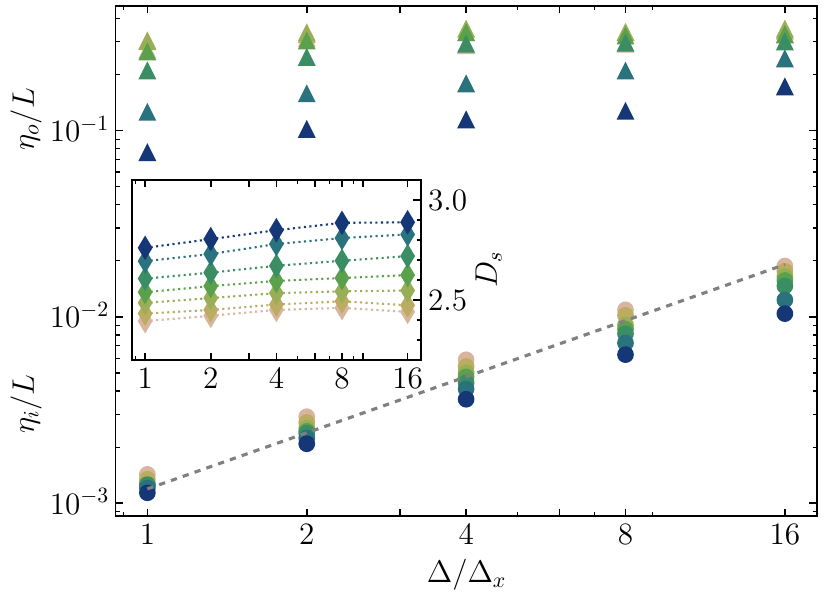}  
  \caption{Same as Fig. \ref{fig:etai_etao}, but using data from different simulations at different resolutions $\Delta_x$} \label{fig:etai_etao_native}
\end{figure}

The evolution of the inner and outer cutoff together with the fractal dimension for different resolution $\Delta_x$ are presented in Fig. \ref{fig:etai_etao_native}. We observe again that the fractal dimension $D_s$ depends mainly on the iso-density threshold, while the influence of $\Delta_x$ is weaker, though measurable. The inner cutoff $\eta_i$ appears linear throughout the range of $\Delta_x$. This means that the cutoff is set by numerical dissipation which increases with $\Delta_x$. Finally, and the most surprising is that the outer cutoff $\eta_o$ which was found to be constant with respect to the filter size now slightly increases with the grid size. We do not have yet an explanation for this, but we note however that the increase is quite weak.

\begin{figure}
  \includegraphics[width=\linewidth]{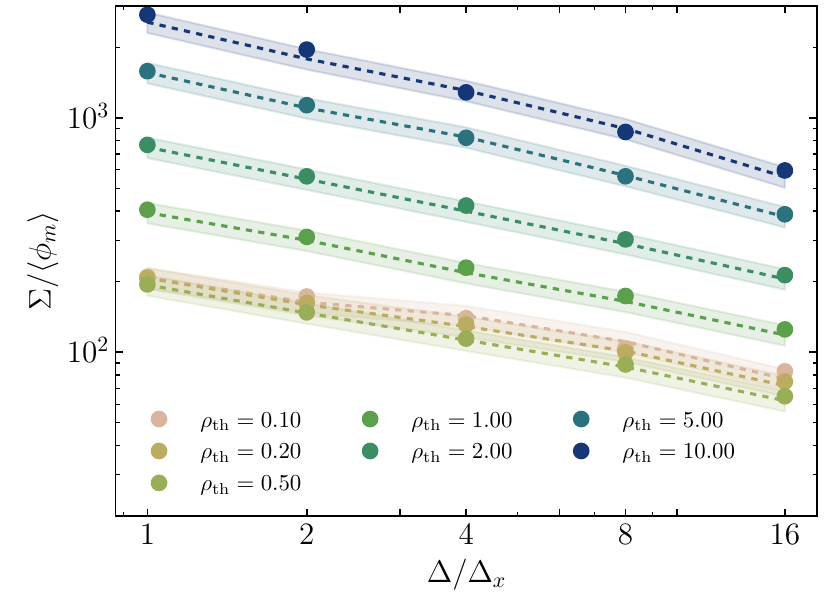}  
  \caption{Same as Fig. \ref{fig:sigma}, but using data from different simulations at different resolutions $\Delta_x$} \label{fig:sigma_native}
\end{figure}

With these values for $\eta_o$, $\eta_i$ and $D_s$, the surface density $\Sigma_s$ can be predicted and compared to the one actually computed. Results are shown in Fig. \ref{fig:sigma_native}. It reveals that the surface-fractal model applies nicely with departures that are within 10\%. The overall conlusion is that, with all other parameters kept unchanged, carrying out a simulation at a given resolution $\Delta_x$ is not equivalent to coarse-graining the finer simulation using a filter size $\Delta_x$.

\bibliographystyle{aa}
\bibliography{astro_geom}

\begin{thebibliography}{91}
\expandafter\ifx\csname natexlab\endcsname\relax\def\natexlab#1{#1}\fi

\bibitem[{Adler {et~al.}(1990)Adler, Jacquin, \& Quiblier}]{Adler1990}
Adler, P.~M., Jacquin, C.~G., \& Quiblier, J.~A. 1990, Int. J. Multiphase Flow,
  16, 691

\bibitem[{Aluie(2013)}]{Aluie_2013}
Aluie, H. 2013, Physica D: Nonlinear Phenomena, 247, 54

\bibitem[{{Appel} {et~al.}(2022){Appel}, {Burkhart}, {Semenov}, {Federrath}, \&
  {Rosen}}]{AppelEtAl2022}
{Appel}, S.~M., {Burkhart}, B., {Semenov}, V.~A., {Federrath}, C., \& {Rosen},
  A.~L. 2022, The Astrophysical Journal, 927, 75

\bibitem[{Audit \& Hennebelle(2010)}]{Audit2010}
Audit, E. \& Hennebelle, P. 2010, Astronomy \& Astrophysics, 511, A76

\bibitem[{Ballesteros-Paredes(2006)}]{Ballesteros-Paredes2006}
Ballesteros-Paredes, J. 2006, Monthly Notices of the Royal Astronomical
  Society, 372, 443

\bibitem[{Ballesteros-Paredes {et~al.}(1999)Ballesteros-Paredes,
  V{\'a}zquez-Semadeni, \& Scalo}]{Ballesteros-Paredes1999}
Ballesteros-Paredes, J., V{\'a}zquez-Semadeni, E., \& Scalo, J. 1999, The
  Astrophysical Journal, 515, 286

\bibitem[{{Beattie} {et~al.}(2019{\natexlab{a}}){Beattie}, {Federrath}, \&
  {Klessen}}]{BeattieFederrathKlessen2019}
{Beattie}, J.~R., {Federrath}, C., \& {Klessen}, R.~S. 2019{\natexlab{a}},
  Monthly Notices of the Royal Astronomical Society, 487, 2070

\bibitem[{{Beattie} {et~al.}(2019{\natexlab{b}}){Beattie}, {Federrath},
  {Klessen}, \& {Schneider}}]{BeattieEtAl2019b}
{Beattie}, J.~R., {Federrath}, C., {Klessen}, R.~S., \& {Schneider}, N.
  2019{\natexlab{b}}, Monthly Notices of the Royal Astronomical Society, 488,
  2493

\bibitem[{Berryman(1987)}]{Berryman1987}
Berryman, J.~G. 1987, J Math Phys, 28, 244

\bibitem[{{Burkhart} \& {Mocz}(2019)}]{BurkhartMocz2019}
{Burkhart}, B. \& {Mocz}, P. 2019, The Astrophysical Journal, 879, 129

\bibitem[{Candel \& Poinsot(1990)}]{Candel1990}
Candel, S. \& Poinsot, T. 1990, Combust. Sci. Technol., 70, 1

\bibitem[{Chandrasekhar \& Fermi(1953)}]{Chandrasekhar1953}
Chandrasekhar, S. \& Fermi, E. 1953, Astrophysical Journal, 116

\bibitem[{Chappell \& Scalo(2001)}]{Chappell2001}
Chappell, D. \& Scalo, J. 2001, The Astrophysical Journal, 551, 712

\bibitem[{Ciccariello(1995)}]{Ciccariello1995}
Ciccariello, S. 1995, J. Math. Phys., 36, 219

\bibitem[{Danaila {et~al.}(2004)Danaila, Antonia, \& Burattini}]{Danaila2004}
Danaila, L., Antonia, R.~A., \& Burattini, P. 2004, New J. Phys., 6, 128

\bibitem[{de~Silva {et~al.}(2013)de~Silva, Philip, Chauhan, Meneveau, \&
  Marusic}]{Silva2013}
de~Silva, C.~M., Philip, J., Chauhan, K., Meneveau, C., \& Marusic, I. 2013,
  Physical review letters, 111, 044501

\bibitem[{Debye {et~al.}(1957)Debye, Anderson~Jr, \& Brumberger}]{Debye1957}
Debye, P., Anderson~Jr, H.~R., \& Brumberger, H. 1957, J Appl Phys, 28, 679

\bibitem[{Dib {et~al.}(2007)Dib, Kim, V{\'a}zquez-Semadeni, Burkert, \&
  Shadmehri}]{Dib2007}
Dib, S., Kim, J., V{\'a}zquez-Semadeni, E., Burkert, A., \& Shadmehri, M. 2007,
  The Astrophysical Journal, 661, 262

\bibitem[{{Dubey} {et~al.}(2008){Dubey}, {Fisher}, {Graziani}, {Jordan},
  {Lamb}, {Reid}, {Rich}, {Sheeler}, {Townsley}, \& {Weide}}]{DubeyEtAl2008}
{Dubey}, A., {Fisher}, R., {Graziani}, C., {et~al.} 2008, in Astronomical
  Society of the Pacific Conference Series, Vol. 385, Numerical Modeling of
  Space Plasma Flows, ed. N.~V. {Pogorelov}, E.~{Audit}, \& G.~P. {Zank}, 145

\bibitem[{Dubrulle(2019)}]{Dubrulle2019}
Dubrulle, B. 2019, J. Fluid Mech., 867, P1

\bibitem[{Duchon \& Robert(2000)}]{Duchon2000}
Duchon, J. \& Robert, R. 2000, Nonlinearity, 13, 249

\bibitem[{Elmegreen \& Falgarone(1996)}]{Elmegreen1996}
Elmegreen, B.~G. \& Falgarone, E. 1996, The Astrophysical Journal, 471, 816

\bibitem[{Elmegreen \& Scalo(2004)}]{Elmegreen2004}
Elmegreen, B.~G. \& Scalo, J. 2004, Annu. Rev. Astron. Astrophys., 42, 211

\bibitem[{Elsas {et~al.}(2018)Elsas, Szalay, \& Meneveau}]{Elsas2018}
Elsas, J.~H., Szalay, A.~S., \& Meneveau, C. 2018, Journal of Turbulence, 19,
  297

\bibitem[{Federrath(2015)}]{Federrath2015}
Federrath, C. 2015, Monthly Notices of the Royal Astronomical Society, 450,
  4035

\bibitem[{Federrath \& Klessen(2012)}]{Federrath2012}
Federrath, C. \& Klessen, R.~S. 2012, The Astrophysical Journal, 761, 156

\bibitem[{Federrath \& Klessen(2013)}]{Federrath2013}
Federrath, C. \& Klessen, R.~S. 2013, The Astrophysical Journal, 763, 51

\bibitem[{Federrath {et~al.}(2021)Federrath, Klessen, Iapichino, \&
  Beattie}]{Federrath2021}
Federrath, C., Klessen, R.~S., Iapichino, L., \& Beattie, J.~R. 2021, Nature
  Astronomy, 5, 365

\bibitem[{{Federrath} {et~al.}(2008){Federrath}, {Klessen}, \&
  {Schmidt}}]{FederrathKlessenSchmidt2008}
{Federrath}, C., {Klessen}, R.~S., \& {Schmidt}, W. 2008, The Astrophysical
  Journal, 688, L79

\bibitem[{Federrath {et~al.}(2009)Federrath, Klessen, \&
  Schmidt}]{Federrath2009}
Federrath, C., Klessen, R.~S., \& Schmidt, W. 2009, The Astrophysical Journal,
  692, 364

\bibitem[{Federrath {et~al.}(2010)Federrath, Roman-Duval, Klessen, Schmidt, \&
  Mac~Low}]{Federrath2010}
Federrath, C., Roman-Duval, J., Klessen, R., Schmidt, W., \& Mac~Low, M.-M.
  2010, Astronomy \& Astrophysics, 512, A81

\bibitem[{{Federrath} {et~al.}(2022){Federrath}, {Roman-Duval}, {Klessen},
  {Schmidt}, \& {Mac Low}}]{FederrathEtAl2022ascl}
{Federrath}, C., {Roman-Duval}, J., {Klessen}, R.~S., {Schmidt}, W., \& {Mac
  Low}, M.~M. 2022, {TG: Turbulence Generator}, Astrophysics Source Code
  Library, record ascl:2204.001

\bibitem[{Ferrand {et~al.}(2020)Ferrand, Galtier, Sahraoui, \&
  Federrath}]{Ferrand2020}
Ferrand, R., Galtier, S., Sahraoui, F., \& Federrath, C. 2020, The
  Astrophysical Journal, 904, 160

\bibitem[{Frisch \& Stillinger(1963)}]{Frisch1963}
Frisch, H.~L. \& Stillinger, F.~H. 1963, J. Chem. Phys., 38, 2200

\bibitem[{Fryxell {et~al.}(2000)Fryxell, Olson, Ricker, Timmes, Zingale, Lamb,
  MacNeice, Rosner, Truran, \& Tufo}]{Fryxell2000}
Fryxell, B., Olson, K., Ricker, P., {et~al.} 2000, The Astrophysical Journal
  Supplement Series, 131, 273

\bibitem[{Galtier(2018)}]{Galtier2018}
Galtier, S. 2018, J. Phys. A: Math. Theor., 51, 205501

\bibitem[{Galtier \& Banerjee(2011)}]{Galtier2011}
Galtier, S. \& Banerjee, S. 2011, Phys Rev Lett, 107, 134501

\bibitem[{Gauding {et~al.}(2022)Gauding, Thiesset, Varea, \&
  Danaila}]{Gauding2022}
Gauding, M., Thiesset, F., Varea, E., \& Danaila, L. 2022, J. Fluid Mech., 942

\bibitem[{{Girichidis} {et~al.}(2014){Girichidis}, {Konstandin}, {Whitworth},
  \& {Klessen}}]{GirichidisEtAl2014}
{Girichidis}, P., {Konstandin}, L., {Whitworth}, A.~P., \& {Klessen}, R.~S.
  2014, The Astrophysical Journal, 781, 91

\bibitem[{Guinier {et~al.}(1955)Guinier, Fournet, \& Yudowitch}]{Guinier1955}
Guinier, A., Fournet, G., \& Yudowitch, K.~L. 1955, Small-angle scattering of
  X-rays (Wiley New York)

\bibitem[{Hawkes {et~al.}(2012)Hawkes, Chatakonda, Kolla, Kerstein, \&
  Chen}]{Hawkes2012}
Hawkes, E.~R., Chatakonda, O., Kolla, H., Kerstein, A.~R., \& Chen, J.~H. 2012,
  Combustion and flame, 159, 2690

\bibitem[{Hennebelle \& Chabrier(2008)}]{Hennebelle2008}
Hennebelle, P. \& Chabrier, G. 2008, The Astrophysical Journal, 684, 395

\bibitem[{Hennebelle \& Falgarone(2012)}]{Hennebelle2012}
Hennebelle, P. \& Falgarone, E. 2012, The Astronomy and Astrophysics Review,
  20, 1

\bibitem[{Hentschel \& Procaccia(1984)}]{Hentschel1984}
Hentschel, H. G.~E. \& Procaccia, I. 1984, Physical Review A, 29, 1461

\bibitem[{{Heyer} \& {Brunt}(2004)}]{HeyerBrunt2004}
{Heyer}, M.~H. \& {Brunt}, C.~M. 2004, The Astrophysical Journal, 615, L45

\bibitem[{Hopkins(2013)}]{Hopkins2013}
Hopkins, P.~F. 2013, Mon. Not. R. Astron. Soc., 430, 1880

\bibitem[{{Kainulainen} {et~al.}(2014){Kainulainen}, {Federrath}, \&
  {Henning}}]{KainulainenFederrathHenning2014}
{Kainulainen}, J., {Federrath}, C., \& {Henning}, T. 2014, Science, 344, 183

\bibitem[{Khullar {et~al.}(2021)Khullar, Federrath, Krumholz, \&
  Matzner}]{Khullar2021}
Khullar, S., Federrath, C., Krumholz, M.~R., \& Matzner, C.~D. 2021, Monthly
  Notices of the Royal Astronomical Society, 507, 4335

\bibitem[{Kim \& Ryu(2005)}]{Kim2005}
Kim, J. \& Ryu, D. 2005, The Astrophysical Journal, 630, L45

\bibitem[{Kirste \& Porod(1962)}]{Kirste1962}
Kirste, R. \& Porod, G. 1962, Kolloid-Zeitschrift und Zeitschrift f{\"u}r
  Polymere, 184, 1

\bibitem[{Kolmogorov(1941)}]{Kolmogorov1941}
Kolmogorov, A. 1941, Dokl. Akad. Nauk. SSSR, 125, 15

\bibitem[{Kritsuk {et~al.}(2006)Kritsuk, Norman, \& Padoan}]{Kritsuk2006}
Kritsuk, A.~G., Norman, M.~L., \& Padoan, P. 2006, The Astrophysical Journal,
  638, L25

\bibitem[{Kritsuk {et~al.}(2007)Kritsuk, Norman, Padoan, \&
  Wagner}]{Kritsuk2007}
Kritsuk, A.~G., Norman, M.~L., Padoan, P., \& Wagner, R. 2007, The
  Astrophysical Journal, 665, 416

\bibitem[{Kritsuk {et~al.}(2011)Kritsuk, Norman, \& Wagner}]{Kritsuk2011}
Kritsuk, A.~G., Norman, M.~L., \& Wagner, R. 2011, The Astrophysical Journal
  Letters, 727, L20

\bibitem[{Krug {et~al.}(2017)Krug, Holzner, Marusic, \& van
  Reeuwijk}]{Krug2017}
Krug, D., Holzner, M., Marusic, I., \& van Reeuwijk, M. 2017, J. Fluid Mech.,
  820, R3

\bibitem[{Krumholz {et~al.}(2012)Krumholz, Klein, \& McKee}]{Krumholz2012}
Krumholz, M.~R., Klein, R.~I., \& McKee, C.~F. 2012, The Astrophysical Journal,
  754, 71

\bibitem[{{Krumholz} \& {McKee}(2005)}]{Krumholz2005}
{Krumholz}, M.~R. \& {McKee}, C.~F. 2005, The Astrophysical Journal, 630, 250

\bibitem[{Lu \& Tryggvason(2018)}]{Lu2018}
Lu, J. \& Tryggvason, G. 2018, Physical Review Fluids, 3, 084401

\bibitem[{Lu \& Tryggvason(2019)}]{Lu2019}
Lu, J. \& Tryggvason, G. 2019, Physical Review Fluids, 4, 084301

\bibitem[{Mac~Low \& Klessen(2004)}]{MacLow2004}
Mac~Low, M.-M. \& Klessen, R.~S. 2004, Reviews of modern physics, 76, 125

\bibitem[{McKee \& Ostriker(2007)}]{McKee2007}
McKee, C.~F. \& Ostriker, E.~C. 2007, Annu. Rev. Astron. Astrophys., 45, 565

\bibitem[{McKee \& Zweibel(1992)}]{McKee1992}
McKee, C.~F. \& Zweibel, E.~G. 1992, Astrophysical Journal, 399, 551

\bibitem[{Mor{\'a}n {et~al.}(2019)Mor{\'a}n, Fuentes, Liu, \& Yon}]{Moran2019}
Mor{\'a}n, J., Fuentes, A., Liu, F., \& Yon, J. 2019, Computer Physics
  Communications, 239, 225

\bibitem[{Myers {et~al.}(2014)Myers, Klein, Krumholz, \& McKee}]{Myers2014}
Myers, A.~T., Klein, R.~I., Krumholz, M.~R., \& McKee, C.~F. 2014, Monthly
  Notices of the Royal Astronomical Society, 439, 3420

\bibitem[{{Ossenkopf} \& {Mac Low}(2002)}]{OssenkopfMacLow2002}
{Ossenkopf}, V. \& {Mac Low}, M.-M. 2002, Astronomy and Astrophysics, 390, 307

\bibitem[{{Padoan} {et~al.}(2014){Padoan}, {Federrath}, {Chabrier}, {Evans},
  {Johnstone}, {J{\o}rgensen}, {McKee}, \& {Nordlund}}]{PadoanEtAl2014}
{Padoan}, P., {Federrath}, C., {Chabrier}, G., {et~al.} 2014, in Protostars and
  Planets VI, ed. H.~{Beuther}, R.~S. {Klessen}, C.~P. {Dullemond}, \&
  T.~{Henning} (University of Arizona Press), 77--100

\bibitem[{Padoan {et~al.}(2004)Padoan, Jimenez, Juvela, \&
  Nordlund}]{Padoan2004}
Padoan, P., Jimenez, R., Juvela, M., \& Nordlund, {\AA}. 2004, The
  Astrophysical Journal, 604, L49

\bibitem[{Padoan \& Nordlund(2002)}]{Padoan2002}
Padoan, P. \& Nordlund, {\AA}. 2002, The Astrophysical Journal, 576, 870

\bibitem[{Padoan \& Nordlund(2011)}]{Padoan2011}
Padoan, P. \& Nordlund, {\AA}. 2011, The Astrophysical Journal, 730, 40

\bibitem[{Passot \& V{\'a}zquez-Semadeni(1998)}]{Passot1998}
Passot, T. \& V{\'a}zquez-Semadeni, E. 1998, Physical Review E, 58, 4501

\bibitem[{Porod(1951)}]{Porod1951}
Porod, G. 1951, Kolloid-Zeitschrift, 124, 83

\bibitem[{{Roman-Duval} {et~al.}(2010){Roman-Duval}, {Jackson}, {Heyer},
  {Rathborne}, \& {Simon}}]{RomanDuvalEtAl2010}
{Roman-Duval}, J., {Jackson}, J.~M., {Heyer}, M., {Rathborne}, J., \& {Simon},
  R. 2010, The Astrophysical Journal, 723, 492

\bibitem[{{S{\'a}nchez} {et~al.}(2005){S{\'a}nchez}, {Alfaro}, \&
  {P{\'e}rez}}]{SanchezEtAl2005}
{S{\'a}nchez}, N., {Alfaro}, E.~J., \& {P{\'e}rez}, E. 2005, The Astrophysical
  Journal, 625, 849

\bibitem[{Saw {et~al.}(2016)Saw, Kuzzay, Faranda, Guittonneau, Daviaud,
  Wiertel-Gasquet, Padilla, \& Dubrulle}]{Saw2016}
Saw, E.-W., Kuzzay, D., Faranda, D., {et~al.} 2016, Nature communications, 7,
  12466

\bibitem[{{Schneider} {et~al.}(2016){Schneider}, {Bontemps}, {Motte},
  {Ossenkopf}, {Klessen}, {Simon}, {Fechtenbaum}, {Herpin}, {Tremblin},
  {Csengeri}, {Myers}, {Hill}, {Cunningham}, \&
  {Federrath}}]{SchneiderEtAl2016}
{Schneider}, N., {Bontemps}, S., {Motte}, F., {et~al.} 2016, Astronomy and
  Astrophysics, 587, A74

\bibitem[{Sorensen(2001)}]{Sorensen2001}
Sorensen, C.~M. 2001, Aerosol Science \& Technology, 35, 648

\bibitem[{Sreenivasan {et~al.}(1989)Sreenivasan, Ramshankar, \&
  Meneveau}]{Sreenivasan1989}
Sreenivasan, K.~R., Ramshankar, R., \& Meneveau, C. 1989, Proceedings of the
  Royal Society of London. A. Mathematical and Physical Sciences, 421, 79

\bibitem[{Stutzki {et~al.}(1998)Stutzki, Bensch, Heithausen, Ossenkopf, \&
  Zielinsky}]{Stutzki1998}
Stutzki, J., Bensch, F., Heithausen, A., Ossenkopf, V., \& Zielinsky, M. 1998,
  Astronomy and Astrophysics, 336, 697

\bibitem[{Teubner(1990)}]{Teubner1990}
Teubner, M. 1990, The Journal of chemical physics, 92, 4501

\bibitem[{Thiesset {et~al.}(2020)Thiesset, Duret, M{\'e}nard, Dumouchel,
  Reveillon, \& Demoulin}]{Thiesset2020}
Thiesset, F., Duret, B., M{\'e}nard, T., {et~al.} 2020, J. Fluid Mech., 886, A4

\bibitem[{Thiesset {et~al.}(2016)Thiesset, Maurice, Halter, Mazellier,
  Chauveau, \& G{\"o}kalp}]{Thiesset2016}
Thiesset, F., Maurice, G., Halter, F., {et~al.} 2016, Physical Review E, 93,
  013116

\bibitem[{Thiesset {et~al.}(2021)Thiesset, M{\'e}nard, \&
  Dumouchel}]{Thiesset2021}
Thiesset, F., M{\'e}nard, T., \& Dumouchel, C. 2021, J. Fluid Mech., 912, A39

\bibitem[{Thiesset \& Poux(2020)}]{Thiesset2020a}
Thiesset, F. \& Poux, A. 2020, Numerical assessment of the two-point
  statistical equations in liquid/gas flows, Tech. rep., CNRS, Normandy Univ.,
  UNIROUEN, INSA Rouen, CORIA.

\bibitem[{Torquato(2002)}]{Torquato2002}
Torquato, S. 2002, Random Heterogeneous Materials. Microstructure and
  Macroscopic Properties (Springer-Verlag New York)

\bibitem[{Vassilicos(1992)}]{Vassilicos1992}
Vassilicos, J.~C. 1992, in Topological aspects of the dynamics of fluids and
  plasmas, ed. H.~K. Moffatt, G.~M. Zaslavsky, P.~Comte, \& M.~Tabor
  (Springer), 427--442

\bibitem[{Vassilicos \& Hunt(1991)}]{Vassilicos1991}
Vassilicos, J.~C. \& Hunt, J. C.~R. 1991, Proceedings of the Royal Society of
  London. Series A: Mathematical and Physical Sciences, 435, 505

\bibitem[{Vassilicos \& Hunt(1996)}]{Vassilicos1996}
Vassilicos, J.~C. \& Hunt, J. C.~R. 1996, in INSTITUTE OF MATHEMATICS AND ITS
  APPLICATIONS CONFERENCE SERIES, Vol.~56, Oxford University Press, 127--154

\bibitem[{Vazquez-Semadeni(1994)}]{Vazquez-Semadeni1994}
Vazquez-Semadeni, E. 1994, Astrophysical Journal, 423, 681

\bibitem[{Waagan {et~al.}(2011)Waagan, Federrath, \& Klingenberg}]{Waagan2011}
Waagan, K., Federrath, C., \& Klingenberg, C. 2011, J. Comput. Phys., 230, 3331

\bibitem[{Wong \& Cao(1992)}]{Wong1992}
Wong, P.-z. \& Cao, Q.-z. 1992, Phys. Rev. B, 45, 7627

\bibitem[{Yaglom(1949)}]{Yaglom1949}
Yaglom, A. 1949, Dokl. Akad. Nauk. SSSR, 69, 743

\end{thebibliography}

\end{document}